\documentclass[aps,pre,amsmath,amssymb,reprint,superscriptaddress]{revtex4-1}
\usepackage{graphicx}
\usepackage{epsf}
\usepackage{bm}
\usepackage{lipsum}

\def\ls{\lesssim}

\newcommand{\diff}{\mathrm{d}}

\begin{document}
\title{
Bridging the gap between molecular dynamics and hydrodynamics in nanoscale
Brownian motions
}

\author{Keisuke Mizuta}
\affiliation{
Division of Chemical Engineering,
Department of Materials Engineering Science, Graduate
School of Engineering Science, Osaka University, Toyonaka, Osaka 560-8531, Japan
}

\author{Yoshiki Ishii}
\affiliation{
Division of Chemical Engineering,
Department of Materials Engineering Science, Graduate
School of Engineering Science, Osaka University, Toyonaka, Osaka 560-8531, Japan
}

\author{Kang Kim}
\email{kk@cheng.es.osaka-u.ac.jp}
\affiliation{
Division of Chemical Engineering,
Department of Materials Engineering Science, Graduate
School of Engineering Science, Osaka University, Toyonaka, Osaka 560-8531, Japan
}

\author{Nobuyuki Matubayasi}
\email{nobuyuki@cheng.es.osaka-u.ac.jp}
\affiliation{
Division of Chemical Engineering,
Department of Materials Engineering Science, Graduate
School of Engineering Science, Osaka University, Toyonaka, Osaka 560-8531, Japan
}
\affiliation{
Elements Strategy Initiative for Catalysts and Batteries, Kyoto
University, Katsura, Kyoto 615-8520, Japan
}

\date{\today}

\begin{abstract}
Through molecular dynamics simulations,
we examined hydrodynamic behavior of the Brownian motion of fullerene
     particles based on molecular interactions.
The solvation free energy and velocity  
 autocorrelation function (VACF) were calculated by using the Lennard--Jones (LJ) and
 Weeks--Chandler--Andersen (WCA) potentials for the solute-solvent and
 solvent-solvent interactions and by changing the size of the fullerene particles.
We also measured the diffusion constant of the fullerene particles and the shear
 viscosity of the host fluid, and then the hydrodynamic radius $a_\mathrm{HD}$
 was quantified from the Stokes--Einstein relation.
The $a_\mathrm{HD}$ value exceeds that of the gyration radius of the fullerene when the
 solvation free energy exhibits largely negative values using the LJ potential.
In contrast, $a_\mathrm{HD}$ becomes comparable to the size of bare
 fullerene,
 when the solvation free energy is positive using the WCA potential.
Furthermore, the VACF of the fullerene particles is directly comparable with
 the analytical expressions utilizing the Navier--Stokes
 equations both in incompressible and compressible forms.
Hydrodynamic long-time tail $t^{-3/2}$ is demonstrated for timescales longer than
the kinematic time of the momentum diffusion over the particles' size.
However, the VACF in shorter timescales deviates from the hydrodynamic
 description, particularly for smaller fullerene particles and for the
 LJ potential.
This occurs even though the
 compressible effect is considered when characterizing the
 decay of VACF around the sound-propagation timescale over
the particles' size.
These results indicate that the nanoscale Brownian
 motion is influenced by the solvation structure around the solute
 particles originating from the molecular interaction.
\end{abstract}

\maketitle

\section{Introduction}

In colloids, the macroscopic solute particles are dispersed in a liquid solvent.
To predict flow behaviors of colloidal dispersions,
not only the motions of the solute particles but also
their coupling with the solvent dynamics must be considered.
Note that the spatial and temporal scales of
solute particles are in orders of magnitudes larger than those of solvent molecules. 
Thus, the dynamics of colloidal dispersions are mostly governed by the
coupling effects of Brownian motion of the colloidal particles and
the hydrodynamics of the solvent~\cite{Russel:1981ef,
Russel:1992vm, Dhont:1996ub, Bian:2016fz}.

This indicates that the 
colloidal system is a typical example of multi-scale
physics, including hierarchical scales, and offers a good target to be solved
in computational science.
In recent years, many coarse-grained methods have been developed utilizing
the scale separation between solute and solvent particles.
These include
stochastic rotation dynamics/multiple-particle collision dynamics
methods~\cite{Malevanets:2000cj, Padding:2004dk, Padding:2005ez,
Padding:2006fo, Gompper:2009ke, Huang:2012ev, Theers:2016km}, the lattice Boltzmann
method~\cite{Ladd:1993gr, Lobaskin:2004hf, Cates:2004kh,
Chatterji:2005ca, Poblete:2014jl}, Stokesian
dynamics~\cite{Ermak:1978ie, Brady:1988cn}, direct numerical
simulations using the immersed boundary method~\cite{Peskin:2002go,
Atzberger:2007eh, Sharma:2004fp},
Fluid Particle Dynamics~\cite{Tanaka:2000gq, Kodama:2004fz,
Tanaka:2006cl, Furukawa:2010ka, Furukawa:2018dt}, and the Smoothed
Profile Method~\cite{Nakayama:2005fv, Kim:2006io, Yamamoto:2007ft,
Nakayama:2008fi, Iwashita:2008cj, Yamamoto:2008hm,
Yamamoto:2009im, Nakayama:2010hl, Tatsumi:2012ba,
Luo:2008hu}.

The aforementioned methods enable consistent simulation of 
fluctuating hydrodynamic descriptions for colloidal particles.
Their reliability has been conventionally tested by
calculating the velocity autocorrelation function (VACF) of a single
solute particle and comparing it with the analytical solution obtained by solving the generalized
Langevin equation, which considers the hydrodynamic memory effect.
It utilizes the general expression for the frequency-dependent hydrodynamic
friction coefficient $\tilde\zeta(\omega)$ of a rigid spherical particle suspended in a viscous fluid,
Here, $\tilde\zeta(\omega)$
is derived from the Navier--Stokes (NS) equation 
(see more details in Section~\ref{sec:hydrodynamics}).
In particular, the hydrodynamic memory effect induces a long-time tail
in the VACF
owing to algebraic decay $t^{-3/2}$, which was
discovered by Alder and Wainwright using molecular dynamics (MD) simulations~\cite{Alder:1970fd}.
Recently, 
the hydrodynamic memory effect was measured through experiments using
a particle tracking technique~\cite{Franosch:2011hh, Jannasch:2011dw,
Huang:2011bb, Kheifets:2014hq, Mo:2018fda}.

However, 
when the solute particle is of nanoscale and is comparable in size to
the solvent particles, the separation of spatial and temporal scales
becomes unclear and 
the validity of the continuum description becomes questionable.
Moreover, the most generally used assumption is the incompressible
condition for the host fluid; however, its validity also becomes unreasonable
because of the sound effect propagating over the molecular length scale.
That is, the particle momentum is transferred by sound waves at short
time intervals and
by vortex formation around the particles at long time intervals.
Therefore, 
the whole aspect of the multi-scale hierarchy must be clarified by
using all-atom MD simulations.
The question that arises is how the molecular interactions are relevant to
the hydrodynamics interactions occurring by sound propagation and 
momentum diffusion via the kinematic viscosity.

MD simulations have been intensively performed for observing the long-time tail in 
the VACF for pure Lennard--Jones (LJ) and Weeks--Chandler--Andersen (WCA)
fluids~\cite{Levesque:1974fw, Erpenbeck:1985hc, McDonough:2001ca,
Dib:2006fw, Lesnicki:2016hz, Han:2018dv, Ignatyuk:2018jl}.
Very recently, the velocity field generated through MD simulations was 
directly compared with that described by the linearized NS equations~\cite{Han:2018dv}.
The validity of the combined Langevin and hydrodynamic model for the
Brownian motion has also been examined using MD
simulations~\cite{Bocquet:1994fr, Bocquet:1997cv, Nuevo:1998he,
OuldKaddour:2000ix, Schmidt:2003fd, Sokolovskii:2006ef, McPhie:2006bn,
OuldKaddour:2007gv, Jung:2017gn}.
Many efforts have been devoted to discussions of the microscopic
origin of the hydrodynamic radius $a_\mathrm{HD}$ and the boundary condition at 
the solute-solvent interface with respect to the Stokes--Einstein (SE)
relation, $D=k_\mathrm{B}T/(c\pi \eta a_\mathrm{HD})$, between the diffusion
constant $D$ of the solute particle and the shear viscosity of the fluid
$\eta$~\cite{Hansen:2013uv}.
Here, $k_\mathrm{B}$ and $T$ are the Boltzmann constant and the
temperature, respectively.
The constant $c$ is determined by stick ($c=6$) or slip
($c=4$) boundary conditions imposed at the particle-fluid interface.
Note that the concepts of the hydrodynamic radius
$a_\mathrm{HD}$ and its
association with the hydrodynamic boundary condition become ambiguous at
molecular scales.
In fact, 
$a_\mathrm{HD}$ is actually ``defined'' by the SE relation and its value
is influenced by the choice of $c$. 
Given that the (macroscopic)
hydrodynamics is implemented with the stick boundary condition, $c = 6$
is a natural choice when the solute size is to be varied continuously
over a wide range by bearing in mind the macroscopic limit.

In Ref.~\cite{Li:2009fv}, Li
Eqs.~(\ref{eq:incompressible_VACF}) +
 (\ref{eq:compressible_correctioin})demonstrated that the van der Waals interaction between
nanoscale LJ clusters and solvent particles plays a crucial
role in determining the hydrodynamic radius $a_\mathrm{HD}$.
In addition, the effect of solvation free energy on the hydrodynamic
transport of the fullerene particles suspended
in a water solvent has been investigated by Morrone \textit{et al.}~\cite{Morrone:2012ba}
Other MD simulation study has been reported for
a system composed of a LJ cluster suspended in LJ fluids~\cite{Chakraborty:2011jd}.
Remarkably, Chakraborty used MD simulations to demonstrate the crossover
of hydrodynamic effects from compressible to incompressible fluids~\cite{Chakraborty:2011jd}.
However, 
the interplay between the hydrodynamic behavior and solvation free energy
has not been thoroughly elucidated yet.

In this study, 
we used MD simulations to comprehensively examine both hydrodynamic and
thermodynamic properties of nanoscale
fullerene particles dispersed in two types of solvents by using the
LJ and WCA potentials.
The contributions of the present study are threefold.
First, 
we analyzed the solvation free energy of a fullerene particle to
investigate how its solvation structure depends on the molecular interaction.
Second,
we quantified the hydrodynamic radius $a_\mathrm{HD}$ from the
diffusion constant and the SE relation by assuming the stick boundary condition.
We examined the impact of intermolecular interactions on the hydrodynamic
radius $a_\mathrm{HD}$ and discussed the results in terms of the solvation free energy.
Third, we investigated the VACF of the fullerene particle to
characterize the hydrodynamic long-time tail.
The sound propagation effect on the VACF is then discussed.
The VACF in MD simulations was compared with the analytic
expressions utilizing the frequency-dependent friction
$\tilde\zeta(\omega)$, which was obtained by solving the NS equation.

The remainder of this paper is organized as follows.
Section~\ref{sec:hydrodynamics} introduces the hydrodynamic model for
the VACF in the Brownian motion.
We explain the MD simulation details in Section~\ref{sec:model}, and present
the numerical results regarding the solvation free energy, 
hydrodynamic radius, and VACF in Section~\ref{sec:results}.
Our conclusions are drawn in Section~\ref{sec:conclusion}, before presenting
an Appendix that provides the numerical results for the VACF in pure LJ and WCA
fluids.

\section{Overview of hydrodynamic descriptions of VACF}
\label{sec:hydrodynamics}

Here, we briefly review the theoretical descriptions of the hydrodynamics
for the VACF of a colloidal particle.
The generalized Langevin equation for a spherical particle with mass
$M$ suspended in a fluid exhibiting fluctuating hydrodynamics has been
analyzed in various studies~\cite{Zwanzig:1970kr, Chow:1972gr,
Chow:1973cm, Hauge:1973df,
Bedeaux:1974cd, Hinch:1975ba, Metiu:1977hd, Espanol:1995vv, Felderhof:2005da}.
Moreover, Bian \textit{et al.} reviewed the recent progress on the
Brownian motion~\cite{Bian:2016fz}.

The equation of motion is written as
\begin{align}
M\frac{\diff \bm{v}}{\diff t} = - \int_0^t \zeta(t-s)\bm{v}(s)\diff s +
 \bm{R}(t), 
\end{align}
where $\zeta(t)$ and $\bm{R}(t)$ represent the memory kernel of the
friction coefficient and the random force acting on the particle, respectively.
$\bm{R}(t)$ satisfies the fluctuation-dissipation theorem $\langle
\bm{R}(t)\cdot\bm{R}(0)\rangle = 3k_\mathrm{B}T\zeta (t)$ with the
zero-mean value $\langle\bm{R}(t)\rangle = \bm{0}$.
Here, $\langle \cdots\rangle$ represents the ensemble average.
The VACF of the particle is defined as $C(t)=\langle \bm{v}(t)\cdot
\bm{v}(0)\rangle$, the time evolution of which is given by 
\begin{align}
M\frac{\diff C(t)}{\diff t} = -\int_0^t \zeta(t-s) C(s)\diff s.
\end{align}
The Laplace transform into the frequency ($\omega$) domain reduces to 
\begin{align}
\tilde C(\omega) = \frac{MC(0)}{-iM\omega +\tilde \zeta(\omega)},
\label{eq:VACF_omega}
\end{align}
where $C(0)=k_\mathrm{B}T/M$ according to the equipartition theorem.
Note that the zero-frequency limit $\tilde C(0)$ corresponds to the
diffusion constant $D=k_\mathrm{B}T/\zeta_0$ with $\zeta_0=\tilde\zeta(0)$.
This is equivalent to the Einstein relation, where $D$ is determined
via the mean square displacement at long times.

For an incompressible fluid, the linearized NS equation
results in the 
Basset--Boussinesq--Oseen equation,
\begin{align}
\bm{F}(t) = -6\pi \eta a \bm{v} - \frac{M_\mathrm{f}}{2} \frac{\diff
 \bm{v}}{\diff t} - 6a^2
 \sqrt{\pi \eta \rho_\mathrm{f}} \int_0^{t} \frac{\diff \bm{v}/\diff t}{\sqrt{t-s}} \diff s,
\label{eq:BBO}
\end{align}
which describes a
force acting on a spherical particle with instantaneous velocity $\bm{v}$ and
acceleration $\diff \bm{v}/\diff t$
in low-Reynolds-number regimes~\cite{Landau:1987hg}.
$\eta$ and $\rho_\mathrm{f}$ denote the shear viscosity and the mass
density of the solvent fluid, respectively.
In addition, $a$ and $M_\mathrm{f}=4\pi a^3\rho_\mathrm{f}/3$ are the particle radius and the
added mass due to the replacement 
of the fluid by the particle, respectively.
That is, 
the particle is considered to move with the mass $M+M_\mathrm{f}/2$
in the incompressible fluid, where the sound is assumed to propagate
with the infinite speed.

According to Eq.~(\ref{eq:BBO}), the frequency-dependent friction coefficient
$\tilde\zeta(\omega)=-\tilde F(\omega)/\tilde v(\omega)$ is expressed as
\begin{align}
\tilde\zeta(\omega)= 6\pi \eta a -i\omega \frac{M_\mathrm{f}}{2}+6\pi
 a^2\sqrt{-i\omega\eta \rho_\mathrm{f}}.
\label{eq:zeta_omega}
\end{align}
Thus, the zero-frequency limit $\zeta_0=6\pi \eta a$ corresponds
to the Stokes drag force, resulting in the SE
formula, $D=k_\mathrm{B}T/(6\pi \eta a)$, under the stick boundary
condition.
The third term, which is proportional to $\sqrt{\omega}$, causes the decay of
$\zeta(t)$ to $t^{-3/2}$, which is the source of the long-time tail in VACF.
The expression of VACF can be 
written through Eqs.~(\ref{eq:VACF_omega}) and
(\ref{eq:zeta_omega}) to
\begin{align}
C^\mathrm{\nu}(t)  =
 \frac{k_\mathrm{B}T}{3M}\frac{2\rho_\mathrm{p}}{3\rho_\mathrm{f}}\frac{1}{3\pi}
\int_0^\infty \frac{e^{-s t/\tau_\nu}s^{1/2}}{1+\sigma_1 s+\sigma_2 s^2}
 \diff s,
\label{eq:incompressible_VACF}
\end{align}
with the fullerene particle mass density $\rho_\mathrm{p}=M/(4\pi a^3/3)$ and the
kinematic time of the momentum diffusion over the particle size
$\tau_\nu=a^2/\nu$~\cite{Paul:1981cc, Chakraborty:2011jd}.
Here, the kinematic viscosity is defined as $\nu=\eta/\rho_\mathrm{f}$.
Factors $\sigma_1$ and $\sigma_2$ are defined as
$\sigma_1=(1/9)(7-4\rho_\mathrm{p}/\rho_\mathrm{f})$ and $\sigma_2 =
(1/9)^2(1+2\rho_\mathrm{p}/\rho_\mathrm{f})^2$, respectively.
The asymptotic behavior of $C^\mathrm{\nu}(t)$ is expressed as 
$C^\mathrm{\nu}(t)\simeq (2k_\mathrm{B}T/3\rho_\mathrm{f})(4\pi \nu t)^{-3/2}$
for long times.
Note that the zero-time value of the VACF becomes
$C^\mathrm{\nu}(0)=k_\mathrm{B}T/(M+M_\mathrm{f}/2)$ owing to the effect of added
mass $M/2$, which deviates from the result of
the equipartition theorem.

To describe the short-time relaxation of the VACF appropriately, 
a correction term is introduced as
\begin{align}
C^\mathrm{c}(t) = \frac{k_\mathrm{B}T}{M}\frac{e^{-\alpha_1
 t/\tau_\mathrm{c}}}{1+2\rho_\mathrm{p}/\rho_\mathrm{f}}\left[\cos\left(\frac{\alpha_2
 t}{\tau_\mathrm{c}}\right)-\frac{\alpha_1}{\alpha_2}\sin\left(\frac{\alpha_2
 t}{\tau_\mathrm{c}}\right)\right],
\label{eq:compressible_correctioin}
\end{align}
where $\tau_\mathrm{c} = a/c$ denotes the sound propagation time over
the particle size with the speed of sound in a compressible
fluid~\cite{Chow:1973cm, Zwanzig:1975hk, Chakraborty:2011jd}.
In addition, $\alpha_1=(1+\rho_\mathrm{f}/2\rho_\mathrm{p})$ and
$\alpha_2=(1-\rho_\mathrm{f}^2/4\rho_\mathrm{p}^2)^{1/2}$.
The initial value of the VACF given by
$C(t)=C^\mathrm{\nu}(t)+C^\mathrm{c}(t)$ eventually recovers the result of the
equipartition theorem,  $C(0)=k_\mathrm{B}T/M$.
In deriving Eq.~(\ref{eq:compressible_correctioin}), 
the time separation as $\epsilon =
\tau_\mathrm{c}/\tau_\nu \ll 1$ was assumed, in which the contribution of sound propagation to
the decay of the VACF is much faster than that of momentum diffusion
owing to the fluid viscosity.
Here, $\epsilon$ represents the non-dimensional factor required to
characterize the fluid incompressibility~\cite{Tatsumi:2012ba}.
This linear combination formula has been examined via stochastic
rotation dynamics~\cite{Padding:2006fo} and MD simulations~\cite{Chakraborty:2011jd}.

Previous studies have also analyzed the frequency-dependent hydrodynamic
friction coefficient, $\tilde\zeta(\omega)$,
in the compressible fluid by using 
the linearized NS equations and the relationship between
the pressure and density fields, $\nabla p=c^2
\nabla\rho_\mathrm{f}$~\cite{Zwanzig:1970kr, Bedeaux:1974cd, Metiu:1977hd, Felderhof:2005da}.
$\tilde \zeta(\omega)$ is expressed as
\begin{align}
\tilde\zeta(\omega)  = \frac{4\pi}{3}\eta a x^2
 \frac{(1+x)(9-9iy - 2y^2)+x^2(1-iy)}{2x^2(1-iy)-(1+x)y^2-x^2y^2}, 
\label{eq:compressible_VACF}
\end{align}
with $x=a(-i\omega \rho_\mathrm{f}/\eta)^{1/2}$ and $y=a\omega/\tilde c$~\cite{Felderhof:2005da}.
Here, the frequency-dependent speed of sound $\tilde c$ is given by
\begin{align}
\tilde c =
 \left[c^2-\frac{i\omega}{\rho_\mathrm{f}}\left(\frac{4}{3}\eta+\eta_\mathrm{v}\right)\right]^{1/2},
\label{eq:sound_speed}
\end{align}
with the bulk viscosity $\eta_\mathrm{v}$.
Equation~(\ref{eq:compressible_VACF}) can be applied to a high-compressibility fluid
exhibiting $\epsilon >1$, where the sound propagation is slower
than the momentum diffusion.
High-compressibility factor $\epsilon>1$ causes a peculiar
``backtracking,'' which corresponds to the negative contribution in the VACF.
This is due to the inversion of the particle at short time scales;
this in turn is induced by
the non-uniform fluid density field around the moving solute particle~\cite{Felderhof:2005da}.
Note that a sufficiently small ratio of 
$\eta_\mathrm{v}/\eta$ is another important factor for a slower sound
propagation because larger bulk viscosity of a fluid
attenuates sound propagation.
Comparisons with the simulated VACF have been made through multi-particle
collision dynamics~\cite{Belushkin:2011hr, Poblete:2014jl} and direct
numerical simulation of fluctuating hydrodynamics~\cite{Tatsumi:2012ba}.
However, to the best of our knowledge, a thorough examination of the
VACF obtained from Eq.~(\ref{eq:compressible_VACF}) has not yet
been performed via MD simulations .

\section{Model and simulation methods}
\label{sec:model}

\begin{table}[t]
\small
  \caption{Radius of gyration of fullerene particle $a$ (in nm), 
 mass density ratio $\rho_\mathrm{f}/\rho_\mathrm{p}$ between
solvent fluid and fullerene particle, and size ratio $L/a$.
}
  \label{table:fullerene}
  \begin{tabular*}{0.45\textwidth}{@{\extracolsep{\fill}}ccccccc}
    \hline
    \hline
   & C$_{20}$ & C$_{60}$ & C$_{180}$ & C$_{240}$ & C$_{320}$ & C$_{540}$\\
    \hline
   $a$ &0.20 &0.34 &0.60 &0.69 &0.80 & 1.04\\
   $\rho_\mathrm{f}/\rho_\mathrm{p}$ & 0.104 & 0.191 & 0.336 & 0.387 & 0.453 & 0.582\\
   $L/a$ & 104.79 & 58.56 &
	       33.18 & 28.86 & 24.89 &
			   19.14\\
    \hline
    \hline
  \end{tabular*}
\end{table}

\begin{table}[t]
\small
  \caption{Physical parameters concerning host fluids for LJ and WCA
 potentials.
$\eta$ denotes shear viscosity (in $10^{-4}$kg m$^{-1}$ s$^{-1}$).
 $\eta_\mathrm{v}$ denotes the
 bulk viscosity (in $10^{-4}$kg m$^{-1}$ s$^{-1}$), and $c$ denotes the speed of sound (in nm/ps).
}
  \label{table:parameters}
  \begin{tabular*}{0.4\textwidth}{@{\extracolsep{\fill}}cccc}
    \hline
    \hline
     & $\eta$ & $\eta_\mathrm{v}$ & $c$\\
    \hline
    LJ & 1.95 & 1.11 & 0.547\\
    WCA & 1.54 & 0.516 & 0.812\\
    \hline
    \hline
  \end{tabular*}
\end{table}

The Gromacs package was used to conduct 
MD simulations for one fullerene particle
suspended in a solvent consisting of $N=160,000$ Ar molecules~\cite{Hess:2008db, Abraham:2015gj}.
This simulation setup was similar to that in a previous MD simulation study~\cite{Ishii:2016bo}.
For the fullerene particles, C$_n$ ($n=20$, $60$, $120$, $240$, $320$, and
$540$) were used.
These fullerenes are good models of spherical particles.
We utilized the geometrical coordinates provided by Tomanek~\cite{Tomanek:2014ba}.
All the C-C distances in the fullerene were constrained with the LINCS algorithm.
The radii of gyration of the fullerenes $a$
are listed in Table~\ref{table:fullerene}.

The interaction is described by the LJ potential,
$U_\mathrm{LJ}(r)=4\epsilon_{\alpha\beta}[(\sigma_{\alpha\beta}/r)^{12}-(\sigma_{\alpha\beta}/r)^{6}]$,
where $r$ is the distance between two atoms and $\alpha, \beta =$ Ar, C.
The Lorentz–-Berthelot combination rule of
$\sigma_{\alpha\beta}=(\sigma_\alpha+\sigma_\beta)/2$ and
$\epsilon_{\alpha\beta}=\sqrt{\epsilon_\alpha\epsilon_\beta}$ was
utilized for the interactions between Ar and C atoms.
Furthermore, the parameters,
$\sigma_\mathrm{Ar}$ = $\sigma_\mathrm{C}$ = 0.34 nm, 
$\epsilon_\mathrm{Ar}/k_\mathrm{B}$ = 117.8 K and
$\epsilon_\mathrm{C}/k_\mathrm{B}$ = 43.3 K were used, and 
the cutoff distance $r_\mathrm{c}$ was chosen as 1.2 nm or 0.382 nm.
The value 1.2 nm corresponds to the conventional value in the LJ potential,
whereas 0.382 nm corresponds to $r_\mathrm{c}=2^{1/6}\sigma_\mathrm{Ar}$.
This generates a purely repulsive potential, which is the so-called WCA
potential,
$U_\mathrm{WCA}(r)=U_\mathrm{LJ}(r) + \epsilon_{\alpha\beta}$ $(r<r_\mathrm{c})$.
In this study, potentials with these two cutoff lengths are referred to
as LJ and WCA potentials, respectively.
Note that the Ar-Cn and Ar-Ar interactions are of the same type; both of
them are chosen from either LJ or WCA potential.

\begin{figure}[t]
\centering
\includegraphics[width=0.35\textwidth]{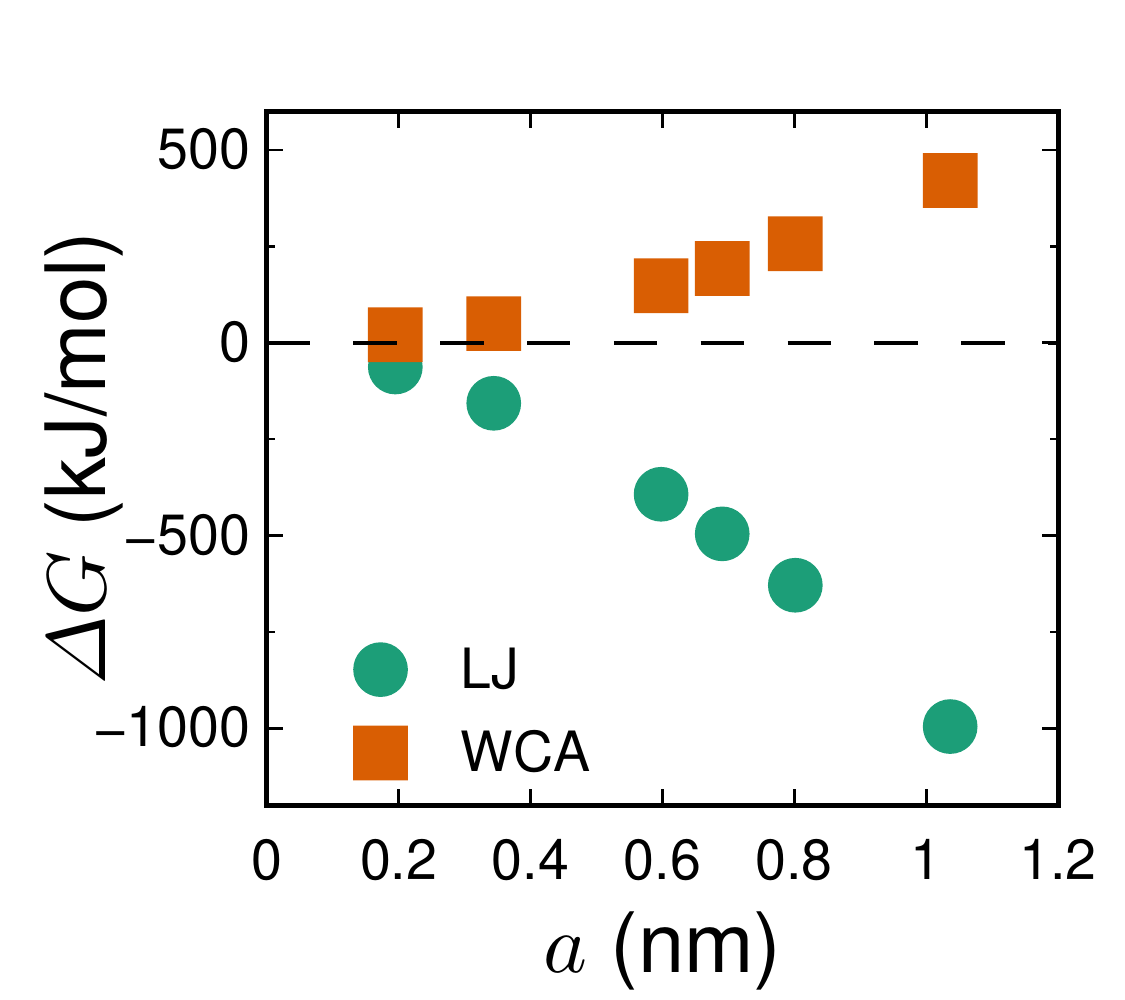}
\caption{
Solvation free energy $\Delta G$ as a function of fullerene particle radius 
for LJ and WCA potential systems.
}
\label{fig:delta_G}
\end{figure}
\begin{figure}[t]
\centering
\includegraphics[width=0.5\textwidth]{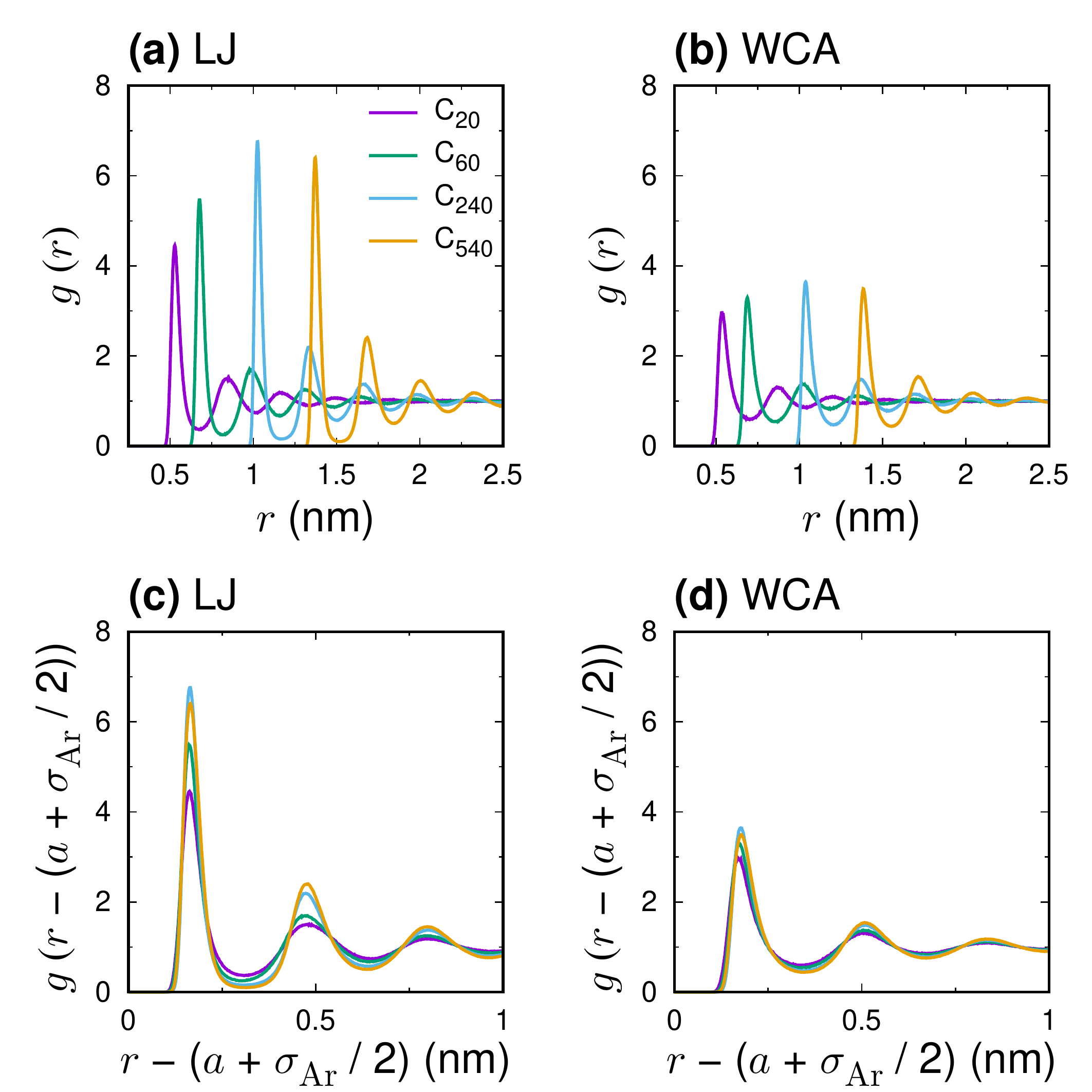}
\caption{
Radial distribution function $g(r)$ for the distance $r$ between center of
 mass of fullerene and solvent particle for LJ (a) and WCA (b) potentials.
Radial distribution function $g(r-(a+\sigma_\mathrm{Ar}/2))$ with the 
 fullerene size $a$ and the Ar radius $\sigma_\mathrm{A}/2$ are also
 plotted for LJ (c) and WCA (d) potentials.
}
\label{fig:rdf}
\end{figure}

The linear dimension of the simulation box was $L=19.91$ nm, and 
the mass and number densities of the solvent were $\rho_\mathrm{N}=20.27$ nm${^{-3}}$ and 
$\rho_\mathrm{f} =\rho_\mathrm{N}m_\mathrm{Ar}= 1,345$ kg/m$^{3}$, respectively.
Here, $m_\mathrm{Ar}$ represents the atomic mass of Ar.
This number density corresponds to $\rho_\mathrm{N}{\sigma_\mathrm{Ar}}^3=0.797$ in
the LJ units.
The mass density of the fullerene is denoted by $\rho_\mathrm{p}=M/(4\pi
a^3/3)$, considering the mass of the fullerene as $M=n m_\mathrm{C}$, where 
$m_\mathrm{C}$ is the atomic mass of carbon.
Then, the mass density ratios between the Ar
solvent fluid and the fullerene particles, \textit{i.e.},
$\rho_\mathrm{f}/\rho_\mathrm{p}$, are presented in Table~\ref{table:fullerene}.
Each system was first equilibrated with the $NVT$ ensemble at the
temperature $T=95$ K, corresponding to
$k_\mathrm{B}T/\epsilon_\mathrm{Ar}=0.806$ in the LJ units.
Then, the $NVE$ ensemble simulations were performed for 10 ns to generate 
20 independent trajectories at each system.
In all simulations,
periodic boundary conditions were utilized with a time step of 1 fs.

The parameters of the host fluid were determined beforehand through MD simulations
for a pure solvent particle system as follows:
Shear viscosity $\eta$ and bulk viscosity $\eta_\mathrm{v}$ were
determined using the Green--Kubo formula for the off-diagonal
and diagonal stress tensor, respectively.
In addition, the speed of sound $c$ was quantified from the numerical
calculations of $\sqrt{(\partial p/\partial \rho_\mathrm{f})_T}$.
The obtained parameters are presented in Table~\ref{table:parameters}.

Periodic images can influence the hydrodynamic behavior in
MD simulations due to its long-range interaction~\cite{Yeh:2004gs}.
The larger size ratio $L/a$ is thus required to characterize the
spatial extent of the momentum transfer with respect to viscosity in MD simulations.
The size ratios $L/a$ of our system are presented in Table~\ref{table:parameters}, 
ranging from 19.14 (C$_{540}$) to 104.79 (C$_{20}$).
Note that MD simulations of LJ cluster dispersions have been performed
to demonstrate long-time tails with the size ratios $L/a=22.58$ and $13.54$ in Ref.~\cite{Chakraborty:2011jd}.
To check the finite-size effects on the VACF, 
we simulated another system with $N=20,000$ Ar particles.
The corresponding linear dimension was $L = 9.96$ nm.
In this smaller system, the desired hydrodynamic long-time tail was
masked by the finite-size artifact and was hardly observed, particularly
in a larger fullerene particle system
(for example, $L/a=9.57$ for C$_{540}$).
Thus, in the following sections, we show the simulation results for $N=160,000$ Ar
particle systems.

\section{Results and discussion}
\label{sec:results}

\begin{figure}[t]
\centering
\includegraphics[width=0.35\textwidth]{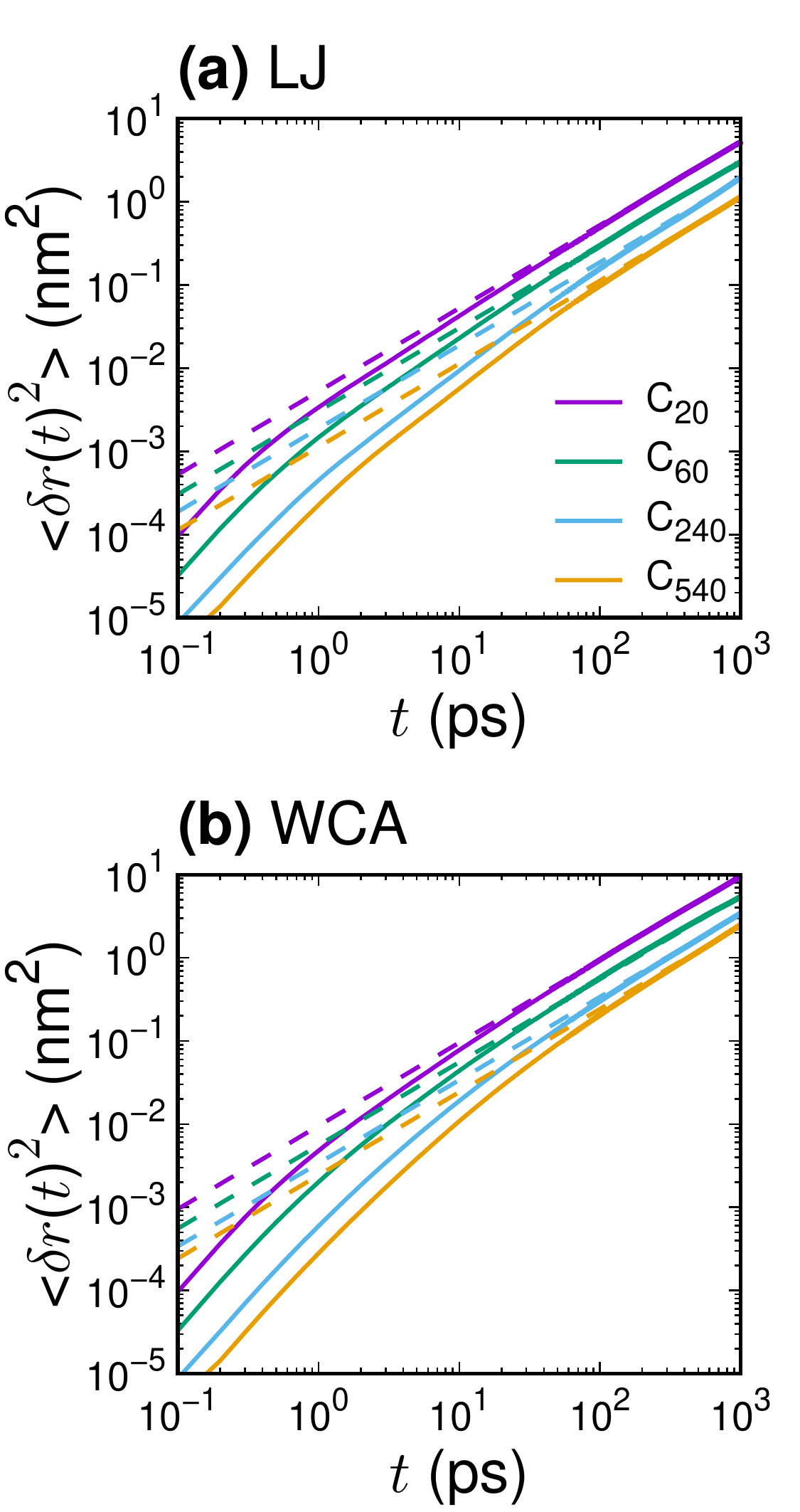}
\caption{
Mean square displacement of the fullerene particle, $\langle \delta r^2(t)\rangle$ for LJ (a) and WCA
 (b) potentials;
dotted lines represent the Einstein relation, 
$\langle \delta r^2(t)\rangle = 6Dt$, using the diffusion constant $D$ for each
 fullerene particle.
}
\label{fig:msd}
\end{figure}

\begin{figure}[t]
\centering
\includegraphics[width=0.35\textwidth]{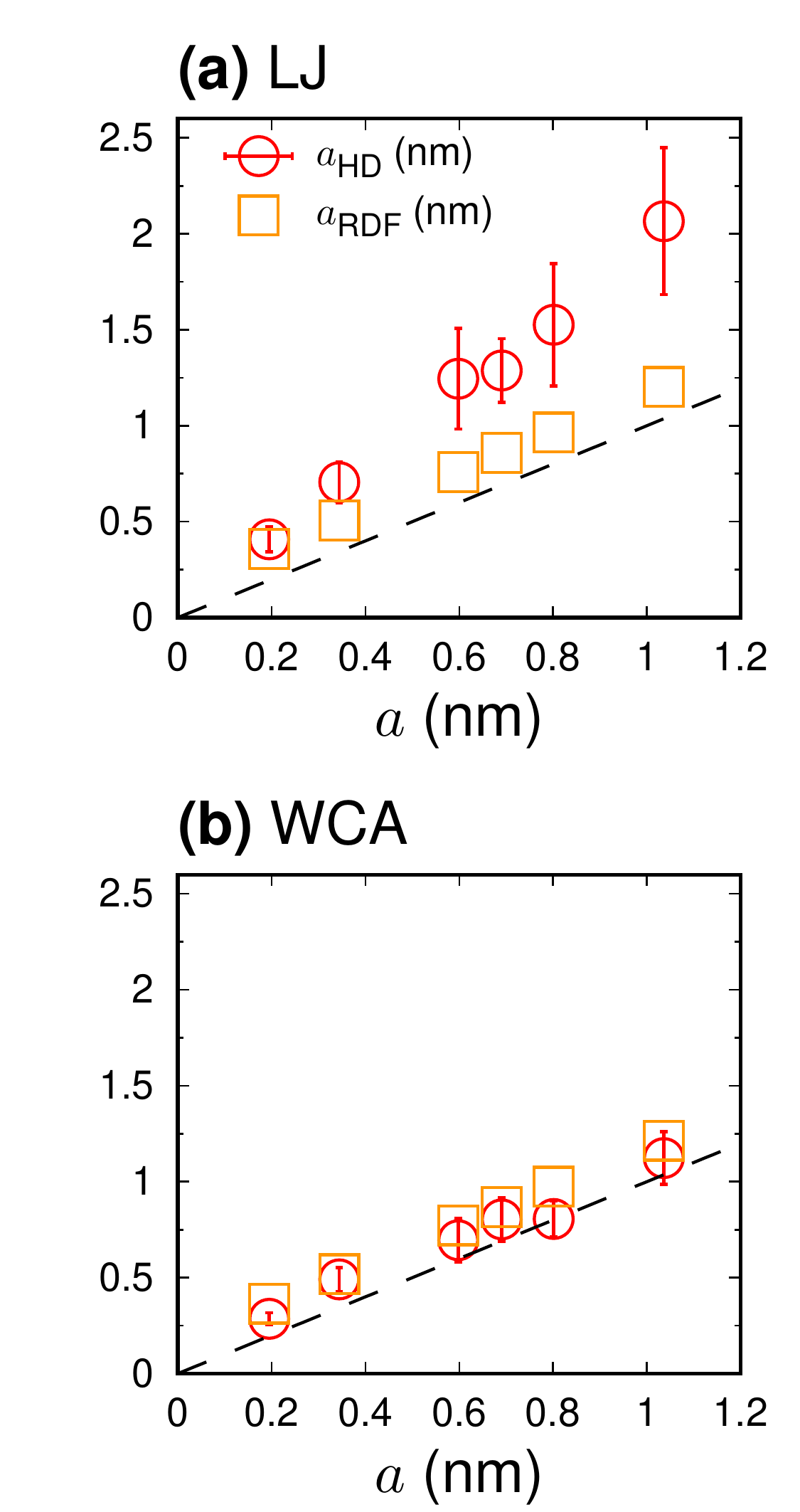}
\caption{
Hydrodynamic radius $a_\mathrm{HD}$ and fullerene size $a_\mathrm{RDF}$
 obtained from the
first solvation shell versus radius of gyration of the
 fullerene particle $a$ for LJ (a) and WCA (b) potentials.
An identical relationship $a_\mathrm{HD}$(or $a_\mathrm{RDF})=a$ is represented by 
 dashed line.
Error bar of $a_\mathrm{HD}$ is represented by standard deviation of data.
}
\label{fig:a_HD}
\end{figure}

\subsection{Solvation free energy}
\label{sec:delta_G}

We first examined the solvation free energy $\Delta G$, which is the
transfer free energy of a solute from a vapor 
to a solvent, for the fullerene particles both in LJ and WCA potential systems.
Note that $\Delta G$ was calculated using the Bennett acceptance ratio
method~\cite{Benett:1976gj}.
Figure~\ref{fig:delta_G} shows the results for $\Delta G$ as a function
of the bare fullerene size, $a$.
$\Delta G$ largely decreases with increasing fullerene size
for the solvent using the LJ potential; this serves as a good solvent
for a larger sized fullerene.
By contrast, the $\Delta G$ of the WCA potential becomes positive,
resulting in a solvation structure that differs from that obtained in the LJ
potential case.
These results lead to the conclusion that the change in solvation free energy of
the fullerene is largely negative owing to the van der Waals 
attraction between the carbons and the solvent Ar molecules.

The solvation structure around the fullerene particle was investigated
with respect to the radial distribution function (RDF), $g(r)$, between
the center of mass of the fullerene
and solvent particles.
The results are shown in Fig.~\ref{fig:rdf}.
As demonstrated in Fig.~\ref{fig:rdf}(a), 
for the solvent using the LJ potential, the maximum peak of $g(r)$
increases with increasing the fullerene size, and correspondingly the
$g(r)$ exhibits the intense oscillation.
This strong solvation structure, in which the fullerene particle is
presumably 
bounded by the solvent particles, is consistent with the negative
$\Delta G$ value resulting from the van der Waals attraction between the fullerene and the
solvent molecules.
In contrast, the peak of $g(r)$ decreases in the case of the
solvent of the WCA potential, as shown in Fig.~\ref{fig:rdf}(b).
This solvophobic property of the fullerene particle also agrees
with the positive value of $\Delta G$.
Furthermore, we examined
$g(r-(a+\sigma_\mathrm{Ar}/2))$ to take into account the peak position
shift with increasing the fullerene size $a$.
The profiles of $g(r-(a+\sigma_\mathrm{Ar}/2))$ are shown in 
Fig.~\ref{fig:rdf}(c) and (d) for the LJ and WCA potentials, respectively.
It is demonstrated that the peak positions are scaled using the distance
$r-(a+\sigma_\mathrm{Ar}/2)$ both in the LJ and WCA systems.
The first maximum positions were observed to be located around $0.17$ nm
in all RDFs.
This size corresponds to $\sigma_\mathrm{C}/2$ because the C atom has a
collision diameter $\sigma_\mathrm{C}$ at the spherical surface with the
gyration radius $a$.
Then, the effective size of the fullerene particle can be defined
from the difference between the first maximum
position $r_\mathrm{max}$ of $g(r)$ and the radius of the solvent
particle, which is expressed as $a_\mathrm{RDF} =r_\mathrm{max} -
\sigma_\mathrm{Ar}/2$.

\subsection{Hydrodynamic radius}

\begin{figure*}[t]
\centering
\includegraphics[width=0.9\textwidth]{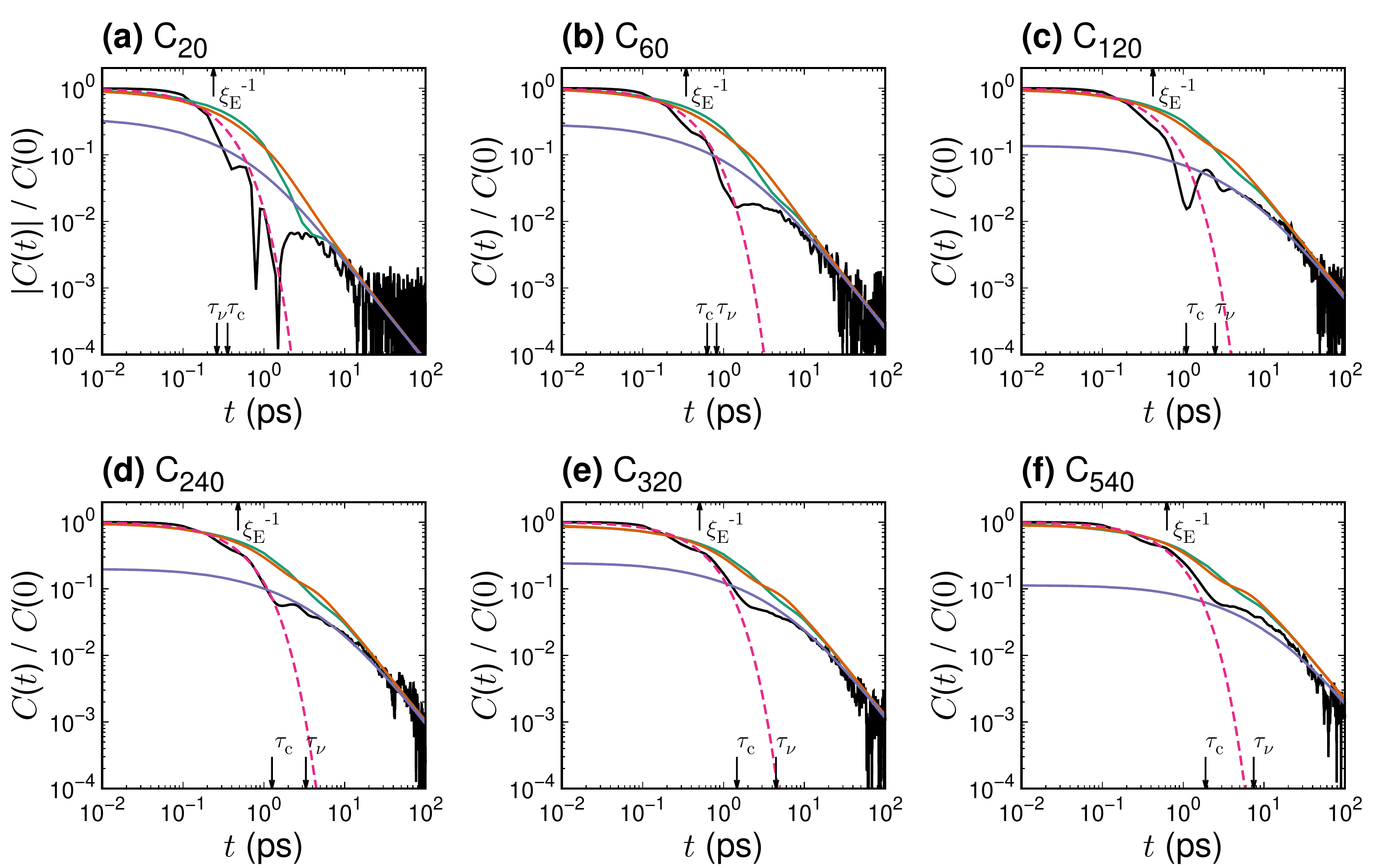}
\caption{
Velocity autocorrelation function $C(t)$ for fullerene particle C$_n$
 of LJ potential with $n=20$ (a), $60$ (b), $120$ (c), $240$ (d), $320$
 (e), and $540$ (f).
Orange and Green curves represent the VACF calculated with
 Eq.~(\ref{eq:compressible_VACF}) (using linearized NS equation for
 compressible fluids) and
combined Eqs.~(\ref{eq:incompressible_VACF}) and
 (\ref{eq:compressible_correctioin}) (using linearized NS equation for
 incompressible fluids along with correction for sound propagation effect), respectively.
Purple curve represents Eq.~(\ref{eq:incompressible_VACF}), where
the hydrodynamic radius $a_\mathrm{HD}$ is utilized for
the solute radius $a$.
Accordingly, the mass density of the solute particle
 $\rho_\mathrm{p}$ was modified as
 $\rho_\mathrm{p}+4\pi\rho_\mathrm{f}(a_\mathrm{HD}^3-a^3)/3$,
 incorporating the mass of the solvent particle within the hydrodynamic
 radius, $a_\mathrm{HD}$.
The dashed magenta line depicts exponential decay with Enskog friction
 coefficient, $\exp(-\zeta_\mathrm{E}t)$.
Kinematic and sound propagation times are indicated as
 $\tau_\nu=a^2/\nu$ and $\tau_\mathrm{c}=a/c$, respectively.
Furthermore, the Enskog time $\xi_\mathrm{E}^{-1}$ is also shown.
}
\label{fig:vacf_LJ}
\end{figure*}

\begin{figure*}[t]
\centering
\includegraphics[width=0.9\textwidth]{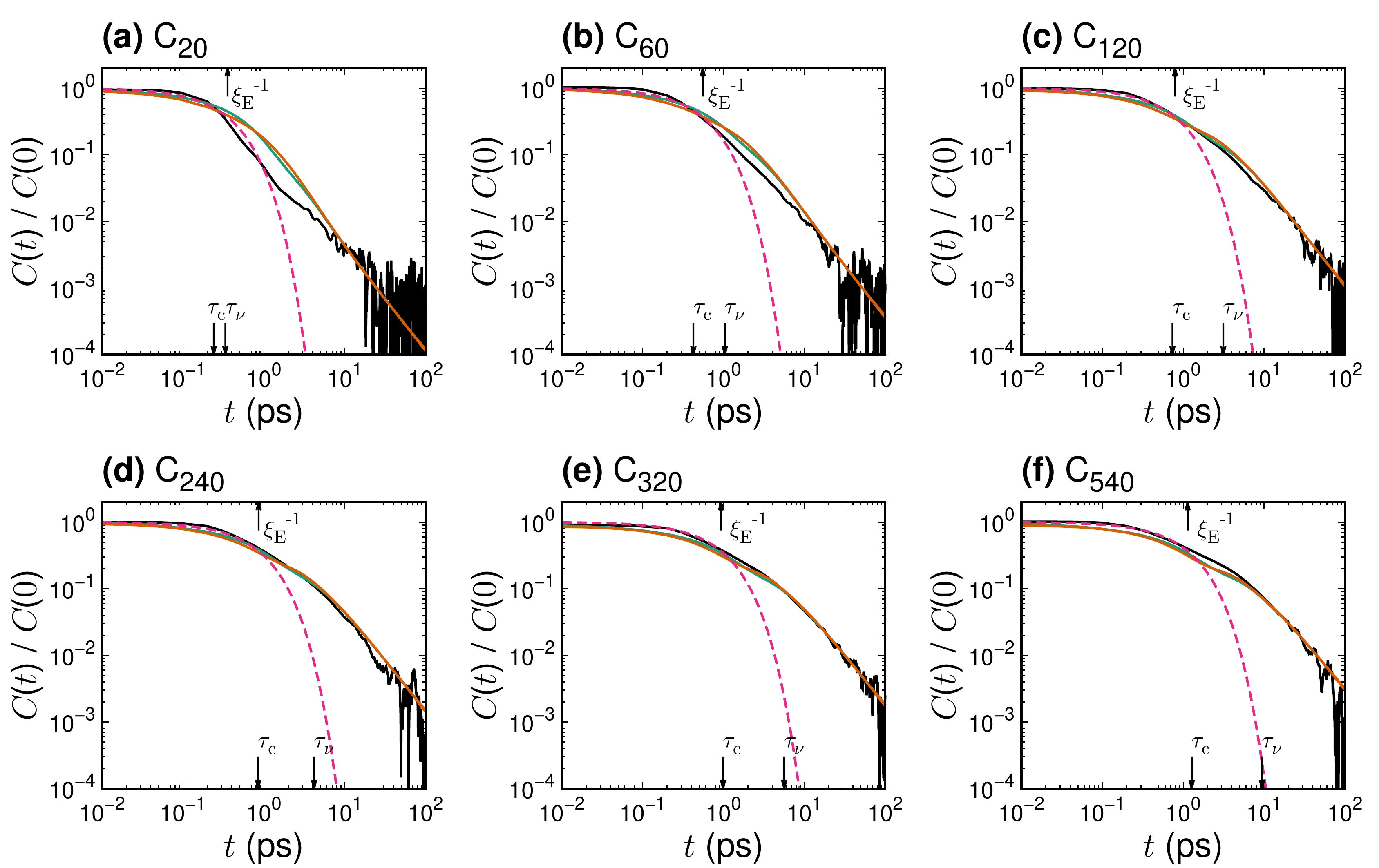}
\caption{
Velocity autocorrelation function $C(t)$ of fullerene particle C$_n$
 of WCA potential with $n=20$ (a), $60$ (b), $120$ (c), $240$ (d), $320$
 (e), and $540$ (f).
Orange, Green and Magenta curves represent quantities identical to those
 described in Fig.~\ref{fig:vacf_LJ}.
Kinematic time and sound propagation times are indicated as
 $\tau_\nu=a^2/\nu$ and $\tau_\mathrm{c}=a/c$, respectively.
Furthermore, the Enskog time $\xi_\mathrm{E}^{-1}$ is also shown.
}
\label{fig:vacf_WCA}
\end{figure*}

\begin{table}[t]
\small
  \caption{Diffusion constant $D$ (in $10^{-4}$nm$^2$/ps) and diffusion
 time $\tau_\mathrm{D}=a^2/D$ (in ps) concerning fullerene particles for LJ and WCA potentials.
}
  \begin{tabular*}{0.45\textwidth}{@{\extracolsep{\fill}}ccccccc}
    \hline
    \hline
     & C$_{20}$ & C$_{60}$ & C$_{180}$ & C$_{240}$ & C$_{320}$ & C$_{540}$\\
    \hline
    $D$ (LJ) & 8.75 & 5.05 & 2.87 & 3.13 & 2.52 & 1.88\\
    $D$ (WCA) & 15.8 & 9.24 & 6.52 & 5.64 & 5.62 & 4.04\\
    $\tau_\mathrm{D}$ (LJ) & 41 & 229 & 1246 & 1522 & 2549 & 5745\\
    $\tau_\mathrm{D}$ (WCA) & 23 & 125 & 548 & 845 & 1143 & 2680\\
    \hline
    \hline
  \end{tabular*}
  \label{table:diffusion}
\end{table}

We calculated the mean square displacement of the fullerene particle,
$\langle \delta r(t)^2\rangle = \langle | \bm{r}(t)-
\bm{r}(0)|^2\rangle$, where $\bm{r}(t)- \bm{r}(0)$ represents the
displacement vector of the center of mass of
the fullerene particle during the time interval $t$.
Diffusion constant $D$ was determined from the Einstein relation,
$D=\lim_{t\to \infty} \langle \delta r(t)^2\rangle/ (6t)$,
as shown in Fig.~\ref{fig:msd}.
Table~\ref{table:diffusion} presents diffusion constant $D$ and 
diffusion time $\tau_\mathrm{D}=a^2/D$, during which the fullerene particle
diffuses over the radius.
Note that the diffusion time $\tau_\mathrm{D}$ is more than an
order of magnitude larger than $\tau_\mathrm{c}$ and $\tau_\nu$.
These results indicate
that the diffusion constant $D$ decreases with increasing $a$,
and correspondingly, diffusion time $\tau_\mathrm{D}$ increases.
Furthermore, 
diffusion constant $D$ seems to reduce owing to the attraction of the LJ
potential compared with the
value obtained using the WCA potential.

Hydrodynamic radius $a_\mathrm{HD}$ was determined through
the SE relation assuming the stick boundary condition of $a_\mathrm{HD} = 
k_\mathrm{B}T/(6\pi D\eta)$.
Figure~\ref{fig:a_HD} shows the
comparison between either $a_\mathrm{HD}$ or $a_\mathrm{RDF}$ and the bare
fullerene size $a$.
We observed that the increasing manner of $a_\mathrm{RDF}$ was akin to
that of $a$ both for the LJ and WCA potentials, by exhibiting the
constant difference $a_\mathrm{RDF}-a \approx \sigma_\mathrm{C}/2$.
These behaviors are consistent with the scaled RDF profiles, 
$g(r-(a+\sigma_\mathrm{Ar}/2))$ (see Fig.~\ref{fig:rdf}(c) and (d)).
However, $a_\mathrm{HD}$ of the LJ potential increases more than 
bare radius $a$, whereas that of the WCA potential is comparable to
$a_\mathrm{RDF}$.
In particular, 
for large fullerene particles in the LJ solvent,
$a_\mathrm{HD}$ was larger than $a_\mathrm{RDF}$ by a
value corresponding to several solvation shells.
Note that the hydrodynamic radius $a_\mathrm{HD}$ will become
a larger value, if we assume the slip boundary condition for the SE relation, $a_\mathrm{HD} = 
k_\mathrm{B}T/(4\pi D\eta)$.

This apparent deviation of $a_\mathrm{HD}$ from $a$ and $a_\mathrm{RDF}$
in the LJ potential, which increases more than $a$,
is explained by the negative value of the solvation free energy $\Delta
G$ and the associated strong solvation structure around the fullerene
particle, as demonstrated in Figs.~\ref{fig:delta_G} and \ref{fig:rdf}.
However, it is reasonable to assume that the hydrodynamic radius
$a_\mathrm{HD}$ will merge into the bare size $a$ at the macroscopic
regime ($a_\mathrm{HD}/a\to 1$), where the size of the solute particle becomes many orders of
magnitude larger than the solvent particle.
This is due to the fact that the spatial resolution
for the molecular size is completely lacked and the 
hydrodynamic description becomes justified 
with the stick boundary condition at the macroscopic regime.

\subsection{VACF}

Numerical results pertaining to the center of mass VACF of 
fullerene particles with regard to LJ and WCA potentials are depicted in
Fig.~\ref{fig:vacf_LJ} and Fig.~\ref{fig:vacf_WCA}, respectively.
Additionally,
numerical results obtained for VACF in pure LJ and WCA fluids are reported in Appendix~\ref{appendix}.

In each plot depicted in Fig.~\ref{fig:vacf_LJ} and
Fig.~\ref{fig:vacf_WCA}, results obtained from MD simulations have been compared against
hydrodynamic descriptions previously explained in Section~\ref{sec:hydrodynamics}.
In addition, the short time decay of the VACF has also been compared against the
Enskog theory, yielding the exponential decay relation,
$C^\mathrm{E}(t)= (k_\mathrm{B}T/M)\exp(-\xi_\mathrm{E} t)$, using the
Enskog friction coefficient given by
\begin{align}
\xi_\mathrm{E} = \frac{8}{3}\left(\frac{2\pi k_\mathrm{B}T
 m_\mathrm{Ar} M}{m_\mathrm{Ar} + M}\right)^{1/2}\frac{\rho_\mathrm{N}
 g(r_\mathrm{max})
 r_\mathrm{max}^2}{m_\mathrm{Ar}}\frac{1+2\chi}{1+\chi}, 
\label{eq:Enskog}
\end{align}
with $\chi = I/Ma^2$ obtained using the moment of inertia $I$ of the
fullerene particles~\cite{Subramanian:1975kv}.
Moreover, $g(r_\mathrm{max})$ denotes 
the peak height of the solute-solvent RDF, $g(r)$, at $r_\mathrm{max}$ (refer Fig.~\ref{fig:rdf}).
It must be noted that the Enskog type exponential decay has a physical
origin different from that of the NS equation.


Figure~\ref{fig:vacf_LJ} demonstrates that
the VACF of all fullerene particles using the LJ potential system exhibits
a long-time tail beyond $\tau_\nu$.
MD simulation results obtained for $t\gg \tau_\nu$ were observed to be
consistent with those obtained using hydrodynamic descriptions of
Eqs.~(\ref{eq:incompressible_VACF}) +
 (\ref{eq:compressible_correctioin}) or
 Eq.~(\ref{eq:compressible_VACF}), the long time asymptote of which 
can be expressed as $(2k_\mathrm{B}T/3\rho_\mathrm{f})(4\pi\nu t)^{-3/2}$.

In the initial time region, VACF results obtained from MD simulations demonstrated
good agreement with analytical hydrodynamics predictions.
This might be puzzling since the VACF is expected to be governed by the Enskog
kinetic theory, yielding the exponential decay,
$C^\mathrm{E}(t)=(k_\mathrm{B}T/M)\exp(-\xi_\mathrm{E} t)$,
owing to limitation pertaining to the continuum description of 
solvent fluids.
Note that the decay time of $C^\mathrm{c}(t)$ can be expressed as
$\tau_\mathrm{c}/(1+\rho_\mathrm{f}/2\rho_\mathrm{p})$ in
Eq.~(\ref{eq:compressible_correctioin}).
From $\rho_\mathrm{f}/\rho_\mathrm{p}$ values in Table~\ref{table:fullerene}, 
this time scale was observed
to be relatively close to the Enskog time $\xi_\mathrm{E}^{-1}$
determined from the MD simulations.
Due to its construction, the hydrodynamic description by
Eqs.~(\ref{eq:incompressible_VACF}) + (\ref{eq:compressible_correctioin})
should agree with the MD results at $t$ close to $0$, and the notable point
is that the short time decay of $C^\mathrm{c}(t)$, \textit{i.e.},
$\tau_\mathrm{c}/(1+\rho_\mathrm{f}/2\rho_\mathrm{p})$,
for the fullerene is close to $\tau_\mathrm{c}$ 
according to the $\rho_\mathrm{f}/\rho_\mathrm{p}$ values in Table~\ref{table:fullerene}.
As shown in Appendix~\ref{appendix}, 
this kind of agreement does not hold in pure LJ and WCA fluids, where
the mass density ratio is estimated as $\rho_\mathrm{f}/\rho_\mathrm{p}
= \rho_\mathrm{N}/(\pi/6) \approx 1.522$.
It has been demonstrated in Fig.~\ref{fig:vacf_LJ} that the Enskog theory
provides a reasonable explanation for short time VACF decays observed
over small time instants, $t\ls \xi_\mathrm{E}^{-1}$.
It must be noted that
VACF of pure LJ and WCA
solvents, for which the tagged solvent particle could be considered as a
consolidated solute,
could be well described using the Enskog theory, as
demonstrated in Appendix~\ref{appendix}.

Deviations from the theoretical expressions described in
Eqs.~(\ref{eq:incompressible_VACF}) +
(\ref{eq:compressible_correctioin}) and Eq.~(\ref{eq:compressible_VACF})
become noticeable during the intermediate time period prior to
commencement of the kinematic time $\tau_\nu$ over which velocity
diffuses the radius of the fullerene particles.
VACF obtained from MD simulations were observed to be less compared to
those obtained from hydrodynamic
descriptions involving linearized NS equation.
This decrease in VACF is directly related to the 
hydrodynamic radius $a_\mathrm{HD}$ which was observed to be larger
compared to the 
bare fullerene radius $a$ in accordance with the following VACF integral, 
\begin{align}
 \frac{k_\mathrm{B}T}{6\pi \eta a_\mathrm{HD}}= \frac{1}{3}\int_0^\infty
 C(t) \diff t.
\end{align}

It is also of interest to observe in Fig.~\ref{fig:vacf_LJ} that
the replacement of $a$ by $a_\mathrm{HD}$ in
Eq.~(\ref{eq:incompressible_VACF}) results in better characterization of 
VACF obtained from MD simulations, particularly with regard to larger fullerene particles,
\textit{e.g.}, C$_n$ ($n \ge 120$).
In this expression, the mass density of the solute particle
correspondingly $\rho_\mathrm{p}$ changes in accordance with the 
relation $\rho_\mathrm{p}+4\pi\rho_\mathrm{f}(a_\mathrm{HD}^3-a^3)/3$,
incorporating mass of the solvent particle within the hydrodynamic
radius, $a_\mathrm{HD}$.
This observation implies that the fullerene particle transport 
occurs in conjunction with that of the surrounding solvation structure,
the size of which is characterized by $a_\mathrm{HD}$.

When the size of the fullerene particles becomes comparable with that of the
solvent particles, the observed value of the compressibility factor
given by $\epsilon=\tau_\mathrm{c}/\tau_\mathrm{\nu}$ increases and finally
exceeds unity in the C$_{20}$ case, as described in Fig.~\ref{fig:vacf_LJ}(a).
As already mentioned in Section.~\ref{sec:hydrodynamics}, high fluid compressibility 
may result in VACF backtracking owing to sound propagation.
In fact, a negative contribution of VACF obtained from MD
simulations can be observed in
Fig.~\ref{fig:vacf_LJ}(a), whereas results of hydrodynamic descriptions
obtained using
Eqs.~(\ref{eq:incompressible_VACF}) + (\ref{eq:compressible_correctioin}) and
Eq.~(\ref{eq:compressible_VACF}) never yield negative VACF values.
In general, the backtracking effect requires a sufficiently small value
of bulk viscosity compared to
shear viscosity~\cite{Felderhof:2005da}.
However, MD simulations provide
finite values of bulk viscosity $\eta_\mathrm{v}$, thereby giving rise to
an attenuation of the sound propagation in accordance with Eq.~(\ref{eq:sound_speed}).

Finally, VACF results obtained for the WCA potential system have been 
plotted in Fig.~\ref{fig:vacf_WCA}, which
demonstrates the 
overall VACF agreement between MD simulations and analytical expressions of
hydrodynamics.
That is, both the
sound propagation effect due to fluid compressibility and long-time tail
caused by
kinematic viscosity can be thoroughly emulated at the molecular
interaction level.
As depicted in Fig.~\ref{fig:a_HD}(b), the hydrodynamic radius
$a_\mathrm{HD}$ approximately equals that of bare fullerene $a$, and
this coincidence is in line with the agreement between MD and
hydrodynamics observed in Fig.~\ref{fig:vacf_WCA}.
However, small deviations were observed for time scales around
$\tau_\nu$, especially with regard to
smaller fullerene particles, C$_{20}$ and C$_{60}$, 
the hydrodynamic radius of which $a_\mathrm{HD}$ slightly exceeds $a$.
Furthermore, over shorter time scales, 
VACF values were observed to be well described by the exponential decay
predicted using the Enskog theory;
this observation agrees well with that corresponding to the LJ potential
system.

\section{Conclusions and Final Remarks}
\label{sec:conclusion}

By performing MD simulations, thermodynamic and hydrodynamic
properties of a single fullerene particle suspended in Ar fluids have
been investigated in this study.
The solvation free energy and the VACF were calculated to
reveal the hydrodynamic behavior of said particles from the viewpoint of the molecular
interaction by using LJ and WCA potentials.

As observed, the solvation free energy $\Delta G$ demonstrated
the strong dependence on the intermolecular potential.
As regards LJ potential,
the attraction energy between fullerene and solvent particles was
observed to 
overwhelms the entropy loss owing to the exclusion of solvent particles,
contributing to
more negative value of $\Delta G$ for larger fullerene particle.
Correspondingly, the solvation was highly structured around the fullerene
particle, as observed in RDF.
In contrast, $\Delta G$ was observed to become positive with
regard to the WCA potential, only utilizing 
the short-range, repulsive part of the LJ potential.

The hydrodynamic radius $a_\mathrm{HD}$ was quantified from the SE
relation using the shear viscosity of the pure solvent and the diffusion
constant of the fullerene particles.
Remarkably, $a_\mathrm{HD}$ of LJ potential was observed to exceed the bare
size of fullerene $a$, whereas the comparable relationship between
$a_\mathrm{HD}$ and $a$ was observed with regard to the WCA potential.
This difference of $a_\mathrm{HD}$ could be attributed to the strength
of solvation quantified by $\Delta G$.
There still exists a difference between $a_\mathrm{HD}$ and $a$ of an
order of a molecular length scale corresponding to several solvation shells.
When the difference $a_\mathrm{HD}-a$ remains at the molecular level,
the ratio $a_\mathrm{HD}/a$
converges to unity for macroscopic values of $a$ (or $a_\mathrm{HD}$),
ensuring that the continuum description remains valid.
We also note the direct evidence of the stick boundary
condition cannot be directly assessed from our MD simulations.
It is still natural to assume the stick boundary condition at the
macroscopic regime, where those hydrodynamics descriptions become valid.
Furthermore, it is speculated that 
$\Delta G$ and $a_\mathrm{HD}$ will depend on the examined thermodynamic
condition by changing density and temperature at molecular scales.

VACF results obtained from MD simulations were directly compared against
those obtained using analytical
expressions based on the generalized Langevin equation and hydrodynamics
involving shear and bulk viscosities as well as the speed of sound of
pure solvents.
As observed, VACF decay demonstrates a long-time tail
$t^{-3/2}$, which is purely governed by the kinematic
viscosity $\nu$ for time scales larger compared to the kinematic time $\tau_\nu$.
For time scales shorter than $\tau_\nu$, the sound propagation effect is
expected to be observed in the VACF.
However, VACF for the LJ potential could not be appropriately predicted using the
hydrodynamic description, albeit the NS equation of compressible fluids
was employed.
In contrast,
VACF for the WCA potential system was observed to be in more accord 
with the corresponding hydrodynamic description
even at approximately the sound propagation time $\tau_\mathrm{c}$,
particularly for larger fullerene particles.
Note that the origin of the difference of the VACF results between MD simulations
and hydrodynamics remains elusive.
To address this concern, it is essential to include not only sound
propagation but also frequency- and wave-number dependent formalism related to
viscoelastic properties of host fluids~\cite{Grimm:2011cu, Puertas:2014ja}.

In summary, the proposed study demonstrated the impact of the intermolecular interaction on
the hydrodynamic behavior in the Brownian motion in all-atom MD simulations.
For a real colloidal particle measuring a radius 1 $\mu$m, it is still
difficult to simulate macroscopic hydrodynamics with molecular descriptions via MD simulations.
In contrast, the proposed simulation system involving nanoscale fullerene particles
enabled to resolve time scales up to the microscopic level.
In particular, results obtained from MD simulations performed in this
study were observed to bridge hierarchical time scales, the Enskog time
$\xi_\mathrm{E}^{-1}$, the sound propagation time $\tau_\mathrm{c}$, and the
kinematic time $\tau_\nu$, and the diffusion time $\tau_\mathrm{D}$.

\begin{acknowledgments}
The authors thank Rei Tatsumi, Takuya Iwashita, Hideyuki Mizuno, and Kazuo Yamada
for helpful discussions.
This work was supported by JSPS KAKENHI Grant Numbers, JP17J01006 (Y.I.),
 JP18H01188 (K.K.), and  JP15K13550 (N.M.).
This work was also supported in part by 
the Post-K Supercomputing Project and the Elements
Strategy Initiative for Catalysts and Batteries from the Ministry of
 Education, Culture, Sports, Science, and Technology.
Y. I. is supported by the JSPS fellowship.
The numerical calculations were performed at Research Center of Computational
Science, Okazaki Research Facilities, National Institutes of Natural Sciences, Japan.
\end{acknowledgments}

\begin{figure}[t]
\centering
\includegraphics[width=0.5\textwidth]{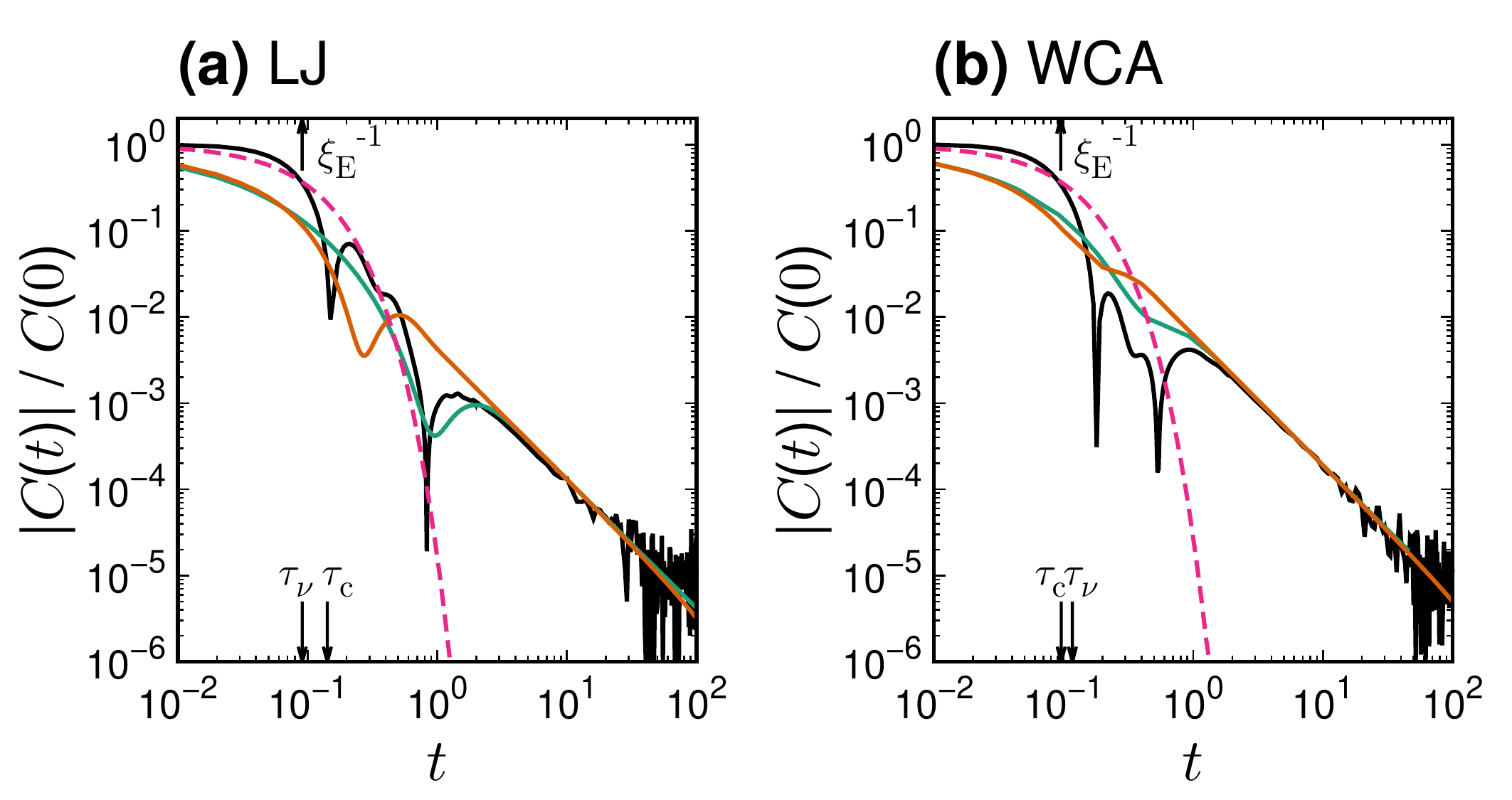}
\caption{
Velocity autocorrelation function $C(t)$ of pure solvent systems using
 LJ (a) and WCA (b) potentials.
Note that the presented quantities are normalized with respect to LJ units for Ar solvent.
Orange, Green and Magenta curves represent quantities identical to those
 described in Fig.~\ref{fig:vacf_LJ}.
Kinematic and sound propagation times are indicated as
 $\tau_\nu=\sigma_\mathrm{Ar}^2/(4\nu)$ and $\tau_\mathrm{c}=\sigma_\mathrm{Ar}/(2c)$,
 respectively.
Enskog time $\xi_\mathrm{E}^{-1}$ is also shown.
}
\label{fig:vacf_solvent}
\end{figure}

\appendix
\section{VACF of pure LJ and WCA fluids}
\label{appendix}

Figure~\ref{fig:vacf_solvent} demonstrates VACF of pure LJ and WCA
fluids using MD simulations.
Similarly to Figs.~\ref{fig:vacf_LJ} and \ref{fig:vacf_WCA}, VACF values
were
compared against those obtained using analytical expressions described in Eqs.~(\ref{eq:incompressible_VACF}) +
 (\ref{eq:compressible_correctioin}) and
 Eq.~(\ref{eq:compressible_VACF}).
Furthermore, values of the exponential decay were plotted in accordance with the Enskog
theory, $\exp(-\xi_\mathrm{E}t)$.
Here,
the Enskog friction coefficient was given by
\begin{align}
\xi_\mathrm{E} = \frac{8}{3}\left(\frac{\pi k_\mathrm{B}T}{m_\mathrm{Ar}}\right)^{1/2} \rho_\mathrm{N}r_\mathrm{max}^2
 g(r_\mathrm{max}),
\end{align}
where $g(r)$ and $r_\mathrm{max}$ represent the RDF and its first peak
position within the system, respectively~\cite{Hansen:2013uv}.
Hydrodynamic radii were quantified as
$a_\mathrm{HD}\approx 0.348\sigma_\mathrm{Ar}$ (LJ) and
$0.346\sigma_\mathrm{Ar}$ (WCA), respectively, values of which were obtained from the SE
relation involving the diffusion constant and shear viscosity.

As demonstrated in Fig.~\ref{fig:vacf_solvent}, the long-time tail is perfectly
characterized through use of the hydrodynamic description, $C(t)\sim
(2k_\mathrm{B}T/3\rho_\mathrm{f})(4\pi\nu t)^{-3/2}$.
This is true for cases involving both LJ and WCA potentials.
However, analytical expressions described in Eqs.~(\ref{eq:incompressible_VACF})
+ (\ref{eq:compressible_correctioin}) and
Eq.~(\ref{eq:compressible_VACF}) demonstrate little ability to reproduce the
MD results over a short time regime.
Alternatively, the Enskog theory demonstrates better agreement with
results obtained using
MD simulations for both LJ and WCA potentials, as illustrated in Fig.~\ref{fig:vacf_solvent}.


\begin{thebibliography}{84}%
\makeatletter
\providecommand \@ifxundefined [1]{%
 \@ifx{#1\undefined}
}%
\providecommand \@ifnum [1]{%
 \ifnum #1\expandafter \@firstoftwo
 \else \expandafter \@secondoftwo
 \fi
}%
\providecommand \@ifx [1]{%
 \ifx #1\expandafter \@firstoftwo
 \else \expandafter \@secondoftwo
 \fi
}%
\providecommand \natexlab [1]{#1}%
\providecommand \enquote  [1]{``#1''}%
\providecommand \bibnamefont  [1]{#1}%
\providecommand \bibfnamefont [1]{#1}%
\providecommand \citenamefont [1]{#1}%
\providecommand \href@noop [0]{\@secondoftwo}%
\providecommand \href [0]{\begingroup \@sanitize@url \@href}%
\providecommand \@href[1]{\@@startlink{#1}\@@href}%
\providecommand \@@href[1]{\endgroup#1\@@endlink}%
\providecommand \@sanitize@url [0]{\catcode `\\12\catcode `\$12\catcode
  `\&12\catcode `\#12\catcode `\^12\catcode `\_12\catcode `\%12\relax}%
\providecommand \@@startlink[1]{}%
\providecommand \@@endlink[0]{}%
\providecommand \url  [0]{\begingroup\@sanitize@url \@url }%
\providecommand \@url [1]{\endgroup\@href {#1}{\urlprefix }}%
\providecommand \urlprefix  [0]{URL }%
\providecommand \Eprint [0]{\href }%
\providecommand \doibase [0]{http://dx.doi.org/}%
\providecommand \selectlanguage [0]{\@gobble}%
\providecommand \bibinfo  [0]{\@secondoftwo}%
\providecommand \bibfield  [0]{\@secondoftwo}%
\providecommand \translation [1]{[#1]}%
\providecommand \BibitemOpen [0]{}%
\providecommand \bibitemStop [0]{}%
\providecommand \bibitemNoStop [0]{.\EOS\space}%
\providecommand \EOS [0]{\spacefactor3000\relax}%
\providecommand \BibitemShut  [1]{\csname bibitem#1\endcsname}%
\let\auto@bib@innerbib\@empty
\bibitem [{\citenamefont {Russel}(1981)}]{Russel:1981ef}%
  \BibitemOpen
  \bibfield  {author} {\bibinfo {author} {\bibfnamefont {W.~B.}\ \bibnamefont
  {Russel}},\ }\href@noop {} {\bibfield  {journal} {\bibinfo  {journal} {Annu.
  Rev. Fluid Mech.}\ }\textbf {\bibinfo {volume} {13}},\ \bibinfo {pages} {425}
  (\bibinfo {year} {1981})}\BibitemShut {NoStop}%
\bibitem [{\citenamefont {Russel}\ \emph {et~al.}(1992)\citenamefont {Russel},
  \citenamefont {Saville},\ and\ \citenamefont {Schowalter}}]{Russel:1992vm}%
  \BibitemOpen
  \bibfield  {author} {\bibinfo {author} {\bibfnamefont {W.~B.}\ \bibnamefont
  {Russel}}, \bibinfo {author} {\bibfnamefont {D.~A.}\ \bibnamefont {Saville}},
  \ and\ \bibinfo {author} {\bibfnamefont {W.~R.}\ \bibnamefont {Schowalter}},\
  }\href@noop {} {\emph {\bibinfo {title} {{Colloidal Dispersions}}}}\
  (\bibinfo  {publisher} {Cambridge University Press},\ \bibinfo {address}
  {Cambridge},\ \bibinfo {year} {1992})\BibitemShut {NoStop}%
\bibitem [{\citenamefont {Dhont}(1996)}]{Dhont:1996ub}%
  \BibitemOpen
  \bibfield  {author} {\bibinfo {author} {\bibfnamefont {J.~K.}\ \bibnamefont
  {Dhont}},\ }\href@noop {} {\emph {\bibinfo {title} {{An introduction to
  dynamics of colloids}}}},\ \bibinfo {series} {Studies in Interface Science},
  Vol.~\bibinfo {volume} {2}\ (\bibinfo  {publisher} {Elsevier},\ \bibinfo
  {address} {Amsterdam},\ \bibinfo {year} {1996})\BibitemShut {NoStop}%
\bibitem [{\citenamefont {Bian}\ \emph {et~al.}(2016)\citenamefont {Bian},
  \citenamefont {Kim},\ and\ \citenamefont {Karniadakis}}]{Bian:2016fz}%
  \BibitemOpen
  \bibfield  {author} {\bibinfo {author} {\bibfnamefont {X.}~\bibnamefont
  {Bian}}, \bibinfo {author} {\bibfnamefont {C.}~\bibnamefont {Kim}}, \ and\
  \bibinfo {author} {\bibfnamefont {G.~E.}\ \bibnamefont {Karniadakis}},\
  }\href@noop {} {\bibfield  {journal} {\bibinfo  {journal} {Soft Matter}\
  }\textbf {\bibinfo {volume} {12}},\ \bibinfo {pages} {6331} (\bibinfo {year}
  {2016})}\BibitemShut {NoStop}%
\bibitem [{\citenamefont {Malevanets}\ and\ \citenamefont
  {Kapral}(2000)}]{Malevanets:2000cj}%
  \BibitemOpen
  \bibfield  {author} {\bibinfo {author} {\bibfnamefont {A.}~\bibnamefont
  {Malevanets}}\ and\ \bibinfo {author} {\bibfnamefont {R.}~\bibnamefont
  {Kapral}},\ }\href@noop {} {\bibfield  {journal} {\bibinfo  {journal} {J.
  Chem. Phys.}\ }\textbf {\bibinfo {volume} {112}},\ \bibinfo {pages} {7260}
  (\bibinfo {year} {2000})}\BibitemShut {NoStop}%
\bibitem [{\citenamefont {Padding}\ and\ \citenamefont
  {Louis}(2004)}]{Padding:2004dk}%
  \BibitemOpen
  \bibfield  {author} {\bibinfo {author} {\bibfnamefont {J.~T.}\ \bibnamefont
  {Padding}}\ and\ \bibinfo {author} {\bibfnamefont {A.~A.}\ \bibnamefont
  {Louis}},\ }\href@noop {} {\bibfield  {journal} {\bibinfo  {journal} {Phys.
  Rev. Lett.}\ }\textbf {\bibinfo {volume} {93}},\ \bibinfo {pages} {220601}
  (\bibinfo {year} {2004})}\BibitemShut {NoStop}%
\bibitem [{\citenamefont {Padding}\ \emph {et~al.}(2005)\citenamefont
  {Padding}, \citenamefont {Wysocki}, \citenamefont {L{\"o}wen},\ and\
  \citenamefont {Louis}}]{Padding:2005ez}%
  \BibitemOpen
  \bibfield  {author} {\bibinfo {author} {\bibfnamefont {J.~T.}\ \bibnamefont
  {Padding}}, \bibinfo {author} {\bibfnamefont {A.}~\bibnamefont {Wysocki}},
  \bibinfo {author} {\bibfnamefont {H.}~\bibnamefont {L{\"o}wen}}, \ and\
  \bibinfo {author} {\bibfnamefont {A.~A.}\ \bibnamefont {Louis}},\ }\href@noop
  {} {\bibfield  {journal} {\bibinfo  {journal} {J. Phys.: Condens. Matter}\
  }\textbf {\bibinfo {volume} {17}},\ \bibinfo {pages} {S3393} (\bibinfo {year}
  {2005})}\BibitemShut {NoStop}%
\bibitem [{\citenamefont {Padding}\ and\ \citenamefont
  {Louis}(2006)}]{Padding:2006fo}%
  \BibitemOpen
  \bibfield  {author} {\bibinfo {author} {\bibfnamefont {J.~T.}\ \bibnamefont
  {Padding}}\ and\ \bibinfo {author} {\bibfnamefont {A.~A.}\ \bibnamefont
  {Louis}},\ }\href@noop {} {\bibfield  {journal} {\bibinfo  {journal} {Phys.
  Rev. E}\ }\textbf {\bibinfo {volume} {74}},\ \bibinfo {pages} {031402}
  (\bibinfo {year} {2006})}\BibitemShut {NoStop}%
\bibitem [{\citenamefont {Gompper}\ \emph {et~al.}(2009)\citenamefont
  {Gompper}, \citenamefont {Ihle}, \citenamefont {Kroll},\ and\ \citenamefont
  {Winkler}}]{Gompper:2009ke}%
  \BibitemOpen
  \bibfield  {author} {\bibinfo {author} {\bibfnamefont {G.}~\bibnamefont
  {Gompper}}, \bibinfo {author} {\bibfnamefont {T.}~\bibnamefont {Ihle}},
  \bibinfo {author} {\bibfnamefont {D.~M.}\ \bibnamefont {Kroll}}, \ and\
  \bibinfo {author} {\bibfnamefont {R.~G.}\ \bibnamefont {Winkler}},\ }in\
  \href@noop {} {\emph {\bibinfo {booktitle} {Advanced Computer Simulation
  Approaches for Soft Matter Sciences III}}}\ (\bibinfo  {publisher}
  {Springer},\ \bibinfo {address} {Berlin, Heidelberg},\ \bibinfo {year}
  {2009})\ pp.\ \bibinfo {pages} {1--87}\BibitemShut {NoStop}%
\bibitem [{\citenamefont {Huang}\ \emph {et~al.}(2012)\citenamefont {Huang},
  \citenamefont {Gompper},\ and\ \citenamefont {Winkler}}]{Huang:2012ev}%
  \BibitemOpen
  \bibfield  {author} {\bibinfo {author} {\bibfnamefont {C.-C.}\ \bibnamefont
  {Huang}}, \bibinfo {author} {\bibfnamefont {G.}~\bibnamefont {Gompper}}, \
  and\ \bibinfo {author} {\bibfnamefont {R.~G.}\ \bibnamefont {Winkler}},\
  }\href@noop {} {\bibfield  {journal} {\bibinfo  {journal} {Phys. Rev. E}\
  }\textbf {\bibinfo {volume} {86}},\ \bibinfo {pages} {056711} (\bibinfo
  {year} {2012})}\BibitemShut {NoStop}%
\bibitem [{\citenamefont {Theers}\ \emph {et~al.}(2016)\citenamefont {Theers},
  \citenamefont {Westphal}, \citenamefont {Gompper},\ and\ \citenamefont
  {Winkler}}]{Theers:2016km}%
  \BibitemOpen
  \bibfield  {author} {\bibinfo {author} {\bibfnamefont {M.}~\bibnamefont
  {Theers}}, \bibinfo {author} {\bibfnamefont {E.}~\bibnamefont {Westphal}},
  \bibinfo {author} {\bibfnamefont {G.}~\bibnamefont {Gompper}}, \ and\
  \bibinfo {author} {\bibfnamefont {R.~G.}\ \bibnamefont {Winkler}},\
  }\href@noop {} {\bibfield  {journal} {\bibinfo  {journal} {Phys. Rev. E}\
  }\textbf {\bibinfo {volume} {93}},\ \bibinfo {pages} {032604} (\bibinfo
  {year} {2016})}\BibitemShut {NoStop}%
\bibitem [{\citenamefont {Ladd}(1993)}]{Ladd:1993gr}%
  \BibitemOpen
  \bibfield  {author} {\bibinfo {author} {\bibfnamefont {A.~J.~C.}\
  \bibnamefont {Ladd}},\ }\href@noop {} {\bibfield  {journal} {\bibinfo
  {journal} {Phys. Rev. Lett.}\ }\textbf {\bibinfo {volume} {70}},\ \bibinfo
  {pages} {1339} (\bibinfo {year} {1993})}\BibitemShut {NoStop}%
\bibitem [{\citenamefont {Lobaskin}\ and\ \citenamefont
  {Dunweg}(2004)}]{Lobaskin:2004hf}%
  \BibitemOpen
  \bibfield  {author} {\bibinfo {author} {\bibfnamefont {V.}~\bibnamefont
  {Lobaskin}}\ and\ \bibinfo {author} {\bibfnamefont {B.}~\bibnamefont
  {Dunweg}},\ }\href@noop {} {\bibfield  {journal} {\bibinfo  {journal} {New J.
  Phys.}\ }\textbf {\bibinfo {volume} {6}},\ \bibinfo {pages} {54} (\bibinfo
  {year} {2004})}\BibitemShut {NoStop}%
\bibitem [{\citenamefont {Cates}\ \emph {et~al.}(2004)\citenamefont {Cates},
  \citenamefont {Stratford}, \citenamefont {Adhikari}, \citenamefont
  {Stansell}, \citenamefont {Desplat}, \citenamefont {Pagonabarraga},\ and\
  \citenamefont {Wagner}}]{Cates:2004kh}%
  \BibitemOpen
  \bibfield  {author} {\bibinfo {author} {\bibfnamefont {M.~E.}\ \bibnamefont
  {Cates}}, \bibinfo {author} {\bibfnamefont {K.}~\bibnamefont {Stratford}},
  \bibinfo {author} {\bibfnamefont {R.}~\bibnamefont {Adhikari}}, \bibinfo
  {author} {\bibfnamefont {P.}~\bibnamefont {Stansell}}, \bibinfo {author}
  {\bibfnamefont {J.-C.}\ \bibnamefont {Desplat}}, \bibinfo {author}
  {\bibfnamefont {I.}~\bibnamefont {Pagonabarraga}}, \ and\ \bibinfo {author}
  {\bibfnamefont {A.~J.}\ \bibnamefont {Wagner}},\ }\href@noop {} {\bibfield
  {journal} {\bibinfo  {journal} {J. Phys.: Condens. Matter}\ }\textbf
  {\bibinfo {volume} {16}},\ \bibinfo {pages} {S3903} (\bibinfo {year}
  {2004})}\BibitemShut {NoStop}%
\bibitem [{\citenamefont {Chatterji}\ and\ \citenamefont
  {Horbach}(2005)}]{Chatterji:2005ca}%
  \BibitemOpen
  \bibfield  {author} {\bibinfo {author} {\bibfnamefont {A.}~\bibnamefont
  {Chatterji}}\ and\ \bibinfo {author} {\bibfnamefont {J.}~\bibnamefont
  {Horbach}},\ }\href@noop {} {\bibfield  {journal} {\bibinfo  {journal} {J.
  Chem. Phys.}\ }\textbf {\bibinfo {volume} {122}},\ \bibinfo {pages} {184903}
  (\bibinfo {year} {2005})}\BibitemShut {NoStop}%
\bibitem [{\citenamefont {Poblete}\ \emph {et~al.}(2014)\citenamefont
  {Poblete}, \citenamefont {Wysocki}, \citenamefont {Gompper},\ and\
  \citenamefont {Winkler}}]{Poblete:2014jl}%
  \BibitemOpen
  \bibfield  {author} {\bibinfo {author} {\bibfnamefont {S.}~\bibnamefont
  {Poblete}}, \bibinfo {author} {\bibfnamefont {A.}~\bibnamefont {Wysocki}},
  \bibinfo {author} {\bibfnamefont {G.}~\bibnamefont {Gompper}}, \ and\
  \bibinfo {author} {\bibfnamefont {R.~G.}\ \bibnamefont {Winkler}},\
  }\href@noop {} {\bibfield  {journal} {\bibinfo  {journal} {Phys. Rev. E}\
  }\textbf {\bibinfo {volume} {90}},\ \bibinfo {pages} {033314} (\bibinfo
  {year} {2014})}\BibitemShut {NoStop}%
\bibitem [{\citenamefont {Ermak}\ and\ \citenamefont
  {McCammon}(1978)}]{Ermak:1978ie}%
  \BibitemOpen
  \bibfield  {author} {\bibinfo {author} {\bibfnamefont {D.~L.}\ \bibnamefont
  {Ermak}}\ and\ \bibinfo {author} {\bibfnamefont {J.~A.}\ \bibnamefont
  {McCammon}},\ }\href@noop {} {\bibfield  {journal} {\bibinfo  {journal} {J.
  Chem. Phys.}\ }\textbf {\bibinfo {volume} {69}},\ \bibinfo {pages} {1352}
  (\bibinfo {year} {1978})}\BibitemShut {NoStop}%
\bibitem [{\citenamefont {Brady}\ and\ \citenamefont
  {Bossis}(1988)}]{Brady:1988cn}%
  \BibitemOpen
  \bibfield  {author} {\bibinfo {author} {\bibfnamefont {J.~F.}\ \bibnamefont
  {Brady}}\ and\ \bibinfo {author} {\bibfnamefont {G.}~\bibnamefont {Bossis}},\
  }\href@noop {} {\bibfield  {journal} {\bibinfo  {journal} {Annu. Rev. Fluid
  Mech.}\ }\textbf {\bibinfo {volume} {20}},\ \bibinfo {pages} {111} (\bibinfo
  {year} {1988})}\BibitemShut {NoStop}%
\bibitem [{\citenamefont {Peskin}(2002)}]{Peskin:2002go}%
  \BibitemOpen
  \bibfield  {author} {\bibinfo {author} {\bibfnamefont {C.~S.}\ \bibnamefont
  {Peskin}},\ }in\ \href@noop {} {\emph {\bibinfo {booktitle} {Acta
  Numerica}}},\ \bibinfo {editor} {edited by\ \bibinfo {editor} {\bibfnamefont
  {A.}~\bibnamefont {Iserles}}}\ (\bibinfo  {publisher} {Cambridge University
  Press},\ \bibinfo {address} {Cambridge},\ \bibinfo {year} {2002})\ pp.\
  \bibinfo {pages} {479--518}\BibitemShut {NoStop}%
\bibitem [{\citenamefont {Atzberger}\ \emph {et~al.}(2007)\citenamefont
  {Atzberger}, \citenamefont {Kramer},\ and\ \citenamefont
  {Peskin}}]{Atzberger:2007eh}%
  \BibitemOpen
  \bibfield  {author} {\bibinfo {author} {\bibfnamefont {P.~J.}\ \bibnamefont
  {Atzberger}}, \bibinfo {author} {\bibfnamefont {P.~R.}\ \bibnamefont
  {Kramer}}, \ and\ \bibinfo {author} {\bibfnamefont {C.~S.}\ \bibnamefont
  {Peskin}},\ }\href@noop {} {\bibfield  {journal} {\bibinfo  {journal} {J.
  Comput. Phys.}\ }\textbf {\bibinfo {volume} {224}},\ \bibinfo {pages} {1255}
  (\bibinfo {year} {2007})}\BibitemShut {NoStop}%
\bibitem [{\citenamefont {Sharma}\ and\ \citenamefont
  {Patankar}(2004)}]{Sharma:2004fp}%
  \BibitemOpen
  \bibfield  {author} {\bibinfo {author} {\bibfnamefont {N.}~\bibnamefont
  {Sharma}}\ and\ \bibinfo {author} {\bibfnamefont {N.~A.}\ \bibnamefont
  {Patankar}},\ }\href@noop {} {\bibfield  {journal} {\bibinfo  {journal} {J.
  Comput. Phys.}\ }\textbf {\bibinfo {volume} {201}},\ \bibinfo {pages} {466}
  (\bibinfo {year} {2004})}\BibitemShut {NoStop}%
\bibitem [{\citenamefont {Tanaka}\ and\ \citenamefont
  {Araki}(2000)}]{Tanaka:2000gq}%
  \BibitemOpen
  \bibfield  {author} {\bibinfo {author} {\bibfnamefont {H.}~\bibnamefont
  {Tanaka}}\ and\ \bibinfo {author} {\bibfnamefont {T.}~\bibnamefont {Araki}},\
  }\href@noop {} {\bibfield  {journal} {\bibinfo  {journal} {Phys. Rev. Lett.}\
  }\textbf {\bibinfo {volume} {85}},\ \bibinfo {pages} {1338} (\bibinfo {year}
  {2000})}\BibitemShut {NoStop}%
\bibitem [{\citenamefont {Kodama}\ \emph {et~al.}(2004)\citenamefont {Kodama},
  \citenamefont {Takeshita}, \citenamefont {Araki},\ and\ \citenamefont
  {Tanaka}}]{Kodama:2004fz}%
  \BibitemOpen
  \bibfield  {author} {\bibinfo {author} {\bibfnamefont {H.}~\bibnamefont
  {Kodama}}, \bibinfo {author} {\bibfnamefont {K.}~\bibnamefont {Takeshita}},
  \bibinfo {author} {\bibfnamefont {T.}~\bibnamefont {Araki}}, \ and\ \bibinfo
  {author} {\bibfnamefont {H.}~\bibnamefont {Tanaka}},\ }\href@noop {}
  {\bibfield  {journal} {\bibinfo  {journal} {J. Phys.: Condens. Matter}\
  }\textbf {\bibinfo {volume} {16}},\ \bibinfo {pages} {L115} (\bibinfo {year}
  {2004})}\BibitemShut {NoStop}%
\bibitem [{\citenamefont {Tanaka}\ and\ \citenamefont
  {Araki}(2006)}]{Tanaka:2006cl}%
  \BibitemOpen
  \bibfield  {author} {\bibinfo {author} {\bibfnamefont {H.}~\bibnamefont
  {Tanaka}}\ and\ \bibinfo {author} {\bibfnamefont {T.}~\bibnamefont {Araki}},\
  }\href@noop {} {\bibfield  {journal} {\bibinfo  {journal} {Chem. Eng. Sci.}\
  }\textbf {\bibinfo {volume} {61}},\ \bibinfo {pages} {2108} (\bibinfo {year}
  {2006})}\BibitemShut {NoStop}%
\bibitem [{\citenamefont {Furukawa}\ and\ \citenamefont
  {Tanaka}(2010)}]{Furukawa:2010ka}%
  \BibitemOpen
  \bibfield  {author} {\bibinfo {author} {\bibfnamefont {A.}~\bibnamefont
  {Furukawa}}\ and\ \bibinfo {author} {\bibfnamefont {H.}~\bibnamefont
  {Tanaka}},\ }\href@noop {} {\bibfield  {journal} {\bibinfo  {journal} {Phys.
  Rev. Lett.}\ }\textbf {\bibinfo {volume} {104}},\ \bibinfo {pages} {245702}
  (\bibinfo {year} {2010})}\BibitemShut {NoStop}%
\bibitem [{\citenamefont {Furukawa}\ \emph {et~al.}(2018)\citenamefont
  {Furukawa}, \citenamefont {Tateno},\ and\ \citenamefont
  {Tanaka}}]{Furukawa:2018dt}%
  \BibitemOpen
  \bibfield  {author} {\bibinfo {author} {\bibfnamefont {A.}~\bibnamefont
  {Furukawa}}, \bibinfo {author} {\bibfnamefont {M.}~\bibnamefont {Tateno}}, \
  and\ \bibinfo {author} {\bibfnamefont {H.}~\bibnamefont {Tanaka}},\
  }\href@noop {} {\bibfield  {journal} {\bibinfo  {journal} {Soft Matter}\
  }\textbf {\bibinfo {volume} {14}},\ \bibinfo {pages} {3738} (\bibinfo {year}
  {2018})}\BibitemShut {NoStop}%
\bibitem [{\citenamefont {Nakayama}\ and\ \citenamefont
  {Yamamoto}(2005)}]{Nakayama:2005fv}%
  \BibitemOpen
  \bibfield  {author} {\bibinfo {author} {\bibfnamefont {Y.}~\bibnamefont
  {Nakayama}}\ and\ \bibinfo {author} {\bibfnamefont {R.}~\bibnamefont
  {Yamamoto}},\ }\href@noop {} {\bibfield  {journal} {\bibinfo  {journal}
  {Phys. Rev. E}\ }\textbf {\bibinfo {volume} {71}},\ \bibinfo {pages} {036707}
  (\bibinfo {year} {2005})}\BibitemShut {NoStop}%
\bibitem [{\citenamefont {Kim}\ \emph {et~al.}(2006)\citenamefont {Kim},
  \citenamefont {Nakayama},\ and\ \citenamefont {Yamamoto}}]{Kim:2006io}%
  \BibitemOpen
  \bibfield  {author} {\bibinfo {author} {\bibfnamefont {K.}~\bibnamefont
  {Kim}}, \bibinfo {author} {\bibfnamefont {Y.}~\bibnamefont {Nakayama}}, \
  and\ \bibinfo {author} {\bibfnamefont {R.}~\bibnamefont {Yamamoto}},\
  }\href@noop {} {\bibfield  {journal} {\bibinfo  {journal} {Phys. Rev. Lett.}\
  }\textbf {\bibinfo {volume} {96}},\ \bibinfo {pages} {208302} (\bibinfo
  {year} {2006})}\BibitemShut {NoStop}%
\bibitem [{\citenamefont {Yamamoto}\ \emph {et~al.}(2007)\citenamefont
  {Yamamoto}, \citenamefont {Kim},\ and\ \citenamefont
  {Nakayama}}]{Yamamoto:2007ft}%
  \BibitemOpen
  \bibfield  {author} {\bibinfo {author} {\bibfnamefont {R.}~\bibnamefont
  {Yamamoto}}, \bibinfo {author} {\bibfnamefont {K.}~\bibnamefont {Kim}}, \
  and\ \bibinfo {author} {\bibfnamefont {Y.}~\bibnamefont {Nakayama}},\
  }\href@noop {} {\bibfield  {journal} {\bibinfo  {journal} {Colloids and
  Surfaces A}\ }\textbf {\bibinfo {volume} {311}},\ \bibinfo {pages} {42}
  (\bibinfo {year} {2007})}\BibitemShut {NoStop}%
\bibitem [{\citenamefont {Nakayama}\ \emph {et~al.}(2008)\citenamefont
  {Nakayama}, \citenamefont {Kim},\ and\ \citenamefont
  {Yamamoto}}]{Nakayama:2008fi}%
  \BibitemOpen
  \bibfield  {author} {\bibinfo {author} {\bibfnamefont {Y.}~\bibnamefont
  {Nakayama}}, \bibinfo {author} {\bibfnamefont {K.}~\bibnamefont {Kim}}, \
  and\ \bibinfo {author} {\bibfnamefont {R.}~\bibnamefont {Yamamoto}},\
  }\href@noop {} {\bibfield  {journal} {\bibinfo  {journal} {Eur. Phys. J. E}\
  }\textbf {\bibinfo {volume} {26}},\ \bibinfo {pages} {361} (\bibinfo {year}
  {2008})}\BibitemShut {NoStop}%
\bibitem [{\citenamefont {Iwashita}\ \emph {et~al.}(2008)\citenamefont
  {Iwashita}, \citenamefont {Nakayama},\ and\ \citenamefont
  {Yamamoto}}]{Iwashita:2008cj}%
  \BibitemOpen
  \bibfield  {author} {\bibinfo {author} {\bibfnamefont {T.}~\bibnamefont
  {Iwashita}}, \bibinfo {author} {\bibfnamefont {Y.}~\bibnamefont {Nakayama}},
  \ and\ \bibinfo {author} {\bibfnamefont {R.}~\bibnamefont {Yamamoto}},\
  }\href@noop {} {\bibfield  {journal} {\bibinfo  {journal} {J. Phys. Soc.
  Jpn.}\ }\textbf {\bibinfo {volume} {77}},\ \bibinfo {pages} {074007}
  (\bibinfo {year} {2008})}\BibitemShut {NoStop}%
\bibitem [{\citenamefont {Yamamoto}\ \emph {et~al.}(2008)\citenamefont
  {Yamamoto}, \citenamefont {Kim}, \citenamefont {Nakayama}, \citenamefont
  {Miyazaki},\ and\ \citenamefont {Reichman}}]{Yamamoto:2008hm}%
  \BibitemOpen
  \bibfield  {author} {\bibinfo {author} {\bibfnamefont {R.}~\bibnamefont
  {Yamamoto}}, \bibinfo {author} {\bibfnamefont {K.}~\bibnamefont {Kim}},
  \bibinfo {author} {\bibfnamefont {Y.}~\bibnamefont {Nakayama}}, \bibinfo
  {author} {\bibfnamefont {K.}~\bibnamefont {Miyazaki}}, \ and\ \bibinfo
  {author} {\bibfnamefont {D.~R.}\ \bibnamefont {Reichman}},\ }\href@noop {}
  {\bibfield  {journal} {\bibinfo  {journal} {J. Phys. Soc. Jpn.}\ }\textbf
  {\bibinfo {volume} {77}},\ \bibinfo {pages} {084804} (\bibinfo {year}
  {2008})}\BibitemShut {NoStop}%
\bibitem [{\citenamefont {Yamamoto}\ \emph {et~al.}(2009)\citenamefont
  {Yamamoto}, \citenamefont {Nakayama},\ and\ \citenamefont
  {Kim}}]{Yamamoto:2009im}%
  \BibitemOpen
  \bibfield  {author} {\bibinfo {author} {\bibfnamefont {R.}~\bibnamefont
  {Yamamoto}}, \bibinfo {author} {\bibfnamefont {Y.}~\bibnamefont {Nakayama}},
  \ and\ \bibinfo {author} {\bibfnamefont {K.}~\bibnamefont {Kim}},\
  }\href@noop {} {\bibfield  {journal} {\bibinfo  {journal} {Int. J. Mod. Phys.
  C}\ }\textbf {\bibinfo {volume} {20}},\ \bibinfo {pages} {1457} (\bibinfo
  {year} {2009})}\BibitemShut {NoStop}%
\bibitem [{\citenamefont {Nakayama}\ \emph {et~al.}(2010)\citenamefont
  {Nakayama}, \citenamefont {Kim},\ and\ \citenamefont
  {Yamamoto}}]{Nakayama:2010hl}%
  \BibitemOpen
  \bibfield  {author} {\bibinfo {author} {\bibfnamefont {Y.}~\bibnamefont
  {Nakayama}}, \bibinfo {author} {\bibfnamefont {K.}~\bibnamefont {Kim}}, \
  and\ \bibinfo {author} {\bibfnamefont {R.}~\bibnamefont {Yamamoto}},\
  }\href@noop {} {\bibfield  {journal} {\bibinfo  {journal} {Adv. Powder
  Technol.}\ }\textbf {\bibinfo {volume} {21}},\ \bibinfo {pages} {206}
  (\bibinfo {year} {2010})}\BibitemShut {NoStop}%
\bibitem [{\citenamefont {Tatsumi}\ and\ \citenamefont
  {Yamamoto}(2012)}]{Tatsumi:2012ba}%
  \BibitemOpen
  \bibfield  {author} {\bibinfo {author} {\bibfnamefont {R.}~\bibnamefont
  {Tatsumi}}\ and\ \bibinfo {author} {\bibfnamefont {R.}~\bibnamefont
  {Yamamoto}},\ }\href@noop {} {\bibfield  {journal} {\bibinfo  {journal}
  {Phys. Rev. E}\ }\textbf {\bibinfo {volume} {85}},\ \bibinfo {pages} {066704}
  (\bibinfo {year} {2012})}\BibitemShut {NoStop}%
\bibitem [{\citenamefont {Luo}\ \emph {et~al.}(2009)\citenamefont {Luo},
  \citenamefont {Maxey},\ and\ \citenamefont {Karniadakis}}]{Luo:2008hu}%
  \BibitemOpen
  \bibfield  {author} {\bibinfo {author} {\bibfnamefont {X.}~\bibnamefont
  {Luo}}, \bibinfo {author} {\bibfnamefont {M.~R.}\ \bibnamefont {Maxey}}, \
  and\ \bibinfo {author} {\bibfnamefont {G.~E.}\ \bibnamefont {Karniadakis}},\
  }\href@noop {} {\bibfield  {journal} {\bibinfo  {journal} {J. Comput. Phys.}\
  }\textbf {\bibinfo {volume} {228}},\ \bibinfo {pages} {1750} (\bibinfo {year}
  {2009})}\BibitemShut {NoStop}%
\bibitem [{\citenamefont {Alder}\ and\ \citenamefont
  {Wainwright}(1970)}]{Alder:1970fd}%
  \BibitemOpen
  \bibfield  {author} {\bibinfo {author} {\bibfnamefont {B.~J.}\ \bibnamefont
  {Alder}}\ and\ \bibinfo {author} {\bibfnamefont {T.~E.}\ \bibnamefont
  {Wainwright}},\ }\href@noop {} {\bibfield  {journal} {\bibinfo  {journal}
  {Phys. Rev. A}\ }\textbf {\bibinfo {volume} {1}},\ \bibinfo {pages} {18}
  (\bibinfo {year} {1970})}\BibitemShut {NoStop}%
\bibitem [{\citenamefont {Franosch}\ \emph {et~al.}(2011)\citenamefont
  {Franosch}, \citenamefont {Grimm}, \citenamefont {Belushkin}, \citenamefont
  {Mor}, \citenamefont {Foffi}, \citenamefont {Forro},\ and\ \citenamefont
  {Jeney}}]{Franosch:2011hh}%
  \BibitemOpen
  \bibfield  {author} {\bibinfo {author} {\bibfnamefont {T.}~\bibnamefont
  {Franosch}}, \bibinfo {author} {\bibfnamefont {M.}~\bibnamefont {Grimm}},
  \bibinfo {author} {\bibfnamefont {M.}~\bibnamefont {Belushkin}}, \bibinfo
  {author} {\bibfnamefont {F.~M.}\ \bibnamefont {Mor}}, \bibinfo {author}
  {\bibfnamefont {G.}~\bibnamefont {Foffi}}, \bibinfo {author} {\bibfnamefont
  {L.}~\bibnamefont {Forro}}, \ and\ \bibinfo {author} {\bibfnamefont
  {S.}~\bibnamefont {Jeney}},\ }\href@noop {} {\bibfield  {journal} {\bibinfo
  {journal} {Nature}\ }\textbf {\bibinfo {volume} {478}},\ \bibinfo {pages}
  {85} (\bibinfo {year} {2011})}\BibitemShut {NoStop}%
\bibitem [{\citenamefont {Jannasch}\ \emph {et~al.}(2011)\citenamefont
  {Jannasch}, \citenamefont {Mahamdeh},\ and\ \citenamefont
  {Sch{\"a}ffer}}]{Jannasch:2011dw}%
  \BibitemOpen
  \bibfield  {author} {\bibinfo {author} {\bibfnamefont {A.}~\bibnamefont
  {Jannasch}}, \bibinfo {author} {\bibfnamefont {M.}~\bibnamefont {Mahamdeh}},
  \ and\ \bibinfo {author} {\bibfnamefont {E.}~\bibnamefont {Sch{\"a}ffer}},\
  }\href@noop {} {\bibfield  {journal} {\bibinfo  {journal} {Phys. Rev. Lett.}\
  }\textbf {\bibinfo {volume} {107}},\ \bibinfo {pages} {228301} (\bibinfo
  {year} {2011})}\BibitemShut {NoStop}%
\bibitem [{\citenamefont {Huang}\ \emph {et~al.}(2011)\citenamefont {Huang},
  \citenamefont {Chavez}, \citenamefont {Taute}, \citenamefont {Luki{\'c}},
  \citenamefont {Jeney}, \citenamefont {Raizen},\ and\ \citenamefont
  {Florin}}]{Huang:2011bb}%
  \BibitemOpen
  \bibfield  {author} {\bibinfo {author} {\bibfnamefont {R.}~\bibnamefont
  {Huang}}, \bibinfo {author} {\bibfnamefont {I.}~\bibnamefont {Chavez}},
  \bibinfo {author} {\bibfnamefont {K.~M.}\ \bibnamefont {Taute}}, \bibinfo
  {author} {\bibfnamefont {B.}~\bibnamefont {Luki{\'c}}}, \bibinfo {author}
  {\bibfnamefont {S.}~\bibnamefont {Jeney}}, \bibinfo {author} {\bibfnamefont
  {M.~G.}\ \bibnamefont {Raizen}}, \ and\ \bibinfo {author} {\bibfnamefont
  {E.-L.}\ \bibnamefont {Florin}},\ }\href@noop {} {\bibfield  {journal}
  {\bibinfo  {journal} {Nat. Phys.}\ }\textbf {\bibinfo {volume} {7}},\
  \bibinfo {pages} {576} (\bibinfo {year} {2011})}\BibitemShut {NoStop}%
\bibitem [{\citenamefont {Kheifets}\ \emph {et~al.}(2014)\citenamefont
  {Kheifets}, \citenamefont {Simha}, \citenamefont {Melin}, \citenamefont
  {Li},\ and\ \citenamefont {Raizen}}]{Kheifets:2014hq}%
  \BibitemOpen
  \bibfield  {author} {\bibinfo {author} {\bibfnamefont {S.}~\bibnamefont
  {Kheifets}}, \bibinfo {author} {\bibfnamefont {A.}~\bibnamefont {Simha}},
  \bibinfo {author} {\bibfnamefont {K.}~\bibnamefont {Melin}}, \bibinfo
  {author} {\bibfnamefont {T.}~\bibnamefont {Li}}, \ and\ \bibinfo {author}
  {\bibfnamefont {M.~G.}\ \bibnamefont {Raizen}},\ }\href@noop {} {\bibfield
  {journal} {\bibinfo  {journal} {Science}\ }\textbf {\bibinfo {volume}
  {343}},\ \bibinfo {pages} {1493} (\bibinfo {year} {2014})}\BibitemShut
  {NoStop}%
\bibitem [{\citenamefont {Mo}\ and\ \citenamefont {Raizen}(2019)}]{Mo:2018fda}%
  \BibitemOpen
  \bibfield  {author} {\bibinfo {author} {\bibfnamefont {J.}~\bibnamefont
  {Mo}}\ and\ \bibinfo {author} {\bibfnamefont {M.~G.}\ \bibnamefont
  {Raizen}},\ }\href@noop {} {\bibfield  {journal} {\bibinfo  {journal} {Annu.
  Rev. Fluid Mech.}\ }\textbf {\bibinfo {volume} {51}},\ \bibinfo {pages} {403}
  (\bibinfo {year} {2019})}\BibitemShut {NoStop}%
\bibitem [{\citenamefont {Levesque}\ and\ \citenamefont
  {Ashurst}(1974)}]{Levesque:1974fw}%
  \BibitemOpen
  \bibfield  {author} {\bibinfo {author} {\bibfnamefont {D.}~\bibnamefont
  {Levesque}}\ and\ \bibinfo {author} {\bibfnamefont {W.~T.}\ \bibnamefont
  {Ashurst}},\ }\href@noop {} {\bibfield  {journal} {\bibinfo  {journal} {Phys.
  Rev. Lett.}\ }\textbf {\bibinfo {volume} {33}},\ \bibinfo {pages} {277}
  (\bibinfo {year} {1974})}\BibitemShut {NoStop}%
\bibitem [{\citenamefont {Erpenbeck}\ and\ \citenamefont
  {Wood}(1985)}]{Erpenbeck:1985hc}%
  \BibitemOpen
  \bibfield  {author} {\bibinfo {author} {\bibfnamefont {J.~J.}\ \bibnamefont
  {Erpenbeck}}\ and\ \bibinfo {author} {\bibfnamefont {W.~W.}\ \bibnamefont
  {Wood}},\ }\href@noop {} {\bibfield  {journal} {\bibinfo  {journal} {Phys.
  Rev. A}\ }\textbf {\bibinfo {volume} {32}},\ \bibinfo {pages} {412} (\bibinfo
  {year} {1985})}\BibitemShut {NoStop}%
\bibitem [{\citenamefont {McDonough}\ \emph {et~al.}(2001)\citenamefont
  {McDonough}, \citenamefont {Russo},\ and\ \citenamefont
  {Snook}}]{McDonough:2001ca}%
  \BibitemOpen
  \bibfield  {author} {\bibinfo {author} {\bibfnamefont {A.}~\bibnamefont
  {McDonough}}, \bibinfo {author} {\bibfnamefont {S.~P.}\ \bibnamefont
  {Russo}}, \ and\ \bibinfo {author} {\bibfnamefont {I.~K.}\ \bibnamefont
  {Snook}},\ }\href@noop {} {\bibfield  {journal} {\bibinfo  {journal} {Phys.
  Rev. E}\ }\textbf {\bibinfo {volume} {63}},\ \bibinfo {pages} {18} (\bibinfo
  {year} {2001})}\BibitemShut {NoStop}%
\bibitem [{\citenamefont {Dib}\ \emph {et~al.}(2006)\citenamefont {Dib},
  \citenamefont {Ould-Kaddour},\ and\ \citenamefont {Levesque}}]{Dib:2006fw}%
  \BibitemOpen
  \bibfield  {author} {\bibinfo {author} {\bibfnamefont {R.~F.~A.}\
  \bibnamefont {Dib}}, \bibinfo {author} {\bibfnamefont {F.}~\bibnamefont
  {Ould-Kaddour}}, \ and\ \bibinfo {author} {\bibfnamefont {D.}~\bibnamefont
  {Levesque}},\ }\href@noop {} {\bibfield  {journal} {\bibinfo  {journal}
  {Phys. Rev. E}\ }\textbf {\bibinfo {volume} {74}},\ \bibinfo {pages} {7}
  (\bibinfo {year} {2006})}\BibitemShut {NoStop}%
\bibitem [{\citenamefont {Lesnicki}\ \emph {et~al.}(2016)\citenamefont
  {Lesnicki}, \citenamefont {Vuilleumier}, \citenamefont {Carof},\ and\
  \citenamefont {Rotenberg}}]{Lesnicki:2016hz}%
  \BibitemOpen
  \bibfield  {author} {\bibinfo {author} {\bibfnamefont {D.}~\bibnamefont
  {Lesnicki}}, \bibinfo {author} {\bibfnamefont {R.}~\bibnamefont
  {Vuilleumier}}, \bibinfo {author} {\bibfnamefont {A.}~\bibnamefont {Carof}},
  \ and\ \bibinfo {author} {\bibfnamefont {B.}~\bibnamefont {Rotenberg}},\
  }\href@noop {} {\bibfield  {journal} {\bibinfo  {journal} {Phys. Rev. Lett.}\
  }\textbf {\bibinfo {volume} {116}},\ \bibinfo {pages} {147804} (\bibinfo
  {year} {2016})}\BibitemShut {NoStop}%
\bibitem [{\citenamefont {Han}\ \emph {et~al.}(2018)\citenamefont {Han},
  \citenamefont {Kim}, \citenamefont {Talkner}, \citenamefont {Karniadakis},\
  and\ \citenamefont {Lee}}]{Han:2018dv}%
  \BibitemOpen
  \bibfield  {author} {\bibinfo {author} {\bibfnamefont {K.~H.}\ \bibnamefont
  {Han}}, \bibinfo {author} {\bibfnamefont {C.}~\bibnamefont {Kim}}, \bibinfo
  {author} {\bibfnamefont {P.}~\bibnamefont {Talkner}}, \bibinfo {author}
  {\bibfnamefont {G.~E.}\ \bibnamefont {Karniadakis}}, \ and\ \bibinfo {author}
  {\bibfnamefont {E.~K.}\ \bibnamefont {Lee}},\ }\href@noop {} {\bibfield
  {journal} {\bibinfo  {journal} {J. Chem. Phys.}\ }\textbf {\bibinfo {volume}
  {148}},\ \bibinfo {pages} {024506} (\bibinfo {year} {2018})}\BibitemShut
  {NoStop}%
\bibitem [{\citenamefont {Ignatyuk}\ \emph {et~al.}(2018)\citenamefont
  {Ignatyuk}, \citenamefont {Mryglod},\ and\ \citenamefont
  {Bryk}}]{Ignatyuk:2018jl}%
  \BibitemOpen
  \bibfield  {author} {\bibinfo {author} {\bibfnamefont {V.~V.}\ \bibnamefont
  {Ignatyuk}}, \bibinfo {author} {\bibfnamefont {I.~M.}\ \bibnamefont
  {Mryglod}}, \ and\ \bibinfo {author} {\bibfnamefont {T.}~\bibnamefont
  {Bryk}},\ }\href@noop {} {\bibfield  {journal} {\bibinfo  {journal} {J. Chem.
  Phys.}\ }\textbf {\bibinfo {volume} {149}},\ \bibinfo {pages} {054101}
  (\bibinfo {year} {2018})}\BibitemShut {NoStop}%
\bibitem [{\citenamefont {Bocquet}\ and\ \citenamefont
  {Barrat}(1994)}]{Bocquet:1994fr}%
  \BibitemOpen
  \bibfield  {author} {\bibinfo {author} {\bibfnamefont {L.}~\bibnamefont
  {Bocquet}}\ and\ \bibinfo {author} {\bibfnamefont {J.-L.}\ \bibnamefont
  {Barrat}},\ }\href@noop {} {\bibfield  {journal} {\bibinfo  {journal} {Phys.
  Rev. E}\ }\textbf {\bibinfo {volume} {49}},\ \bibinfo {pages} {3079}
  (\bibinfo {year} {1994})}\BibitemShut {NoStop}%
\bibitem [{\citenamefont {Bocquet}\ \emph {et~al.}(1997)\citenamefont
  {Bocquet}, \citenamefont {Hansen},\ and\ \citenamefont
  {Piasecki}}]{Bocquet:1997cv}%
  \BibitemOpen
  \bibfield  {author} {\bibinfo {author} {\bibfnamefont {L.}~\bibnamefont
  {Bocquet}}, \bibinfo {author} {\bibfnamefont {J.-P.}\ \bibnamefont {Hansen}},
  \ and\ \bibinfo {author} {\bibfnamefont {J.}~\bibnamefont {Piasecki}},\
  }\href@noop {} {\bibfield  {journal} {\bibinfo  {journal} {J. Stat. Phys.}\
  }\textbf {\bibinfo {volume} {89}},\ \bibinfo {pages} {321} (\bibinfo {year}
  {1997})}\BibitemShut {NoStop}%
\bibitem [{\citenamefont {Nuevo}\ \emph {et~al.}(1998)\citenamefont {Nuevo},
  \citenamefont {Morales},\ and\ \citenamefont {Heyes}}]{Nuevo:1998he}%
  \BibitemOpen
  \bibfield  {author} {\bibinfo {author} {\bibfnamefont {M.~J.}\ \bibnamefont
  {Nuevo}}, \bibinfo {author} {\bibfnamefont {J.~J.}\ \bibnamefont {Morales}},
  \ and\ \bibinfo {author} {\bibfnamefont {D.~M.}\ \bibnamefont {Heyes}},\
  }\href@noop {} {\bibfield  {journal} {\bibinfo  {journal} {Phys. Rev. E}\
  }\textbf {\bibinfo {volume} {58}},\ \bibinfo {pages} {5845} (\bibinfo {year}
  {1998})}\BibitemShut {NoStop}%
\bibitem [{\citenamefont {Ould-Kaddour}\ and\ \citenamefont
  {Levesque}(2000)}]{OuldKaddour:2000ix}%
  \BibitemOpen
  \bibfield  {author} {\bibinfo {author} {\bibfnamefont {F.}~\bibnamefont
  {Ould-Kaddour}}\ and\ \bibinfo {author} {\bibfnamefont {D.}~\bibnamefont
  {Levesque}},\ }\href@noop {} {\bibfield  {journal} {\bibinfo  {journal}
  {Phys. Rev. E}\ }\textbf {\bibinfo {volume} {63}},\ \bibinfo {pages} {011205}
  (\bibinfo {year} {2000})}\BibitemShut {NoStop}%
\bibitem [{\citenamefont {Schmidt}\ and\ \citenamefont
  {Skinner}(2003)}]{Schmidt:2003fd}%
  \BibitemOpen
  \bibfield  {author} {\bibinfo {author} {\bibfnamefont {J.~R.}\ \bibnamefont
  {Schmidt}}\ and\ \bibinfo {author} {\bibfnamefont {J.~L.}\ \bibnamefont
  {Skinner}},\ }\href@noop {} {\bibfield  {journal} {\bibinfo  {journal} {J.
  Chem. Phys.}\ }\textbf {\bibinfo {volume} {119}},\ \bibinfo {pages} {8062}
  (\bibinfo {year} {2003})}\BibitemShut {NoStop}%
\bibitem [{\citenamefont {Sokolovskii}\ \emph {et~al.}(2006)\citenamefont
  {Sokolovskii}, \citenamefont {Thachuk},\ and\ \citenamefont
  {Patey}}]{Sokolovskii:2006ef}%
  \BibitemOpen
  \bibfield  {author} {\bibinfo {author} {\bibfnamefont {R.~O.}\ \bibnamefont
  {Sokolovskii}}, \bibinfo {author} {\bibfnamefont {M.}~\bibnamefont
  {Thachuk}}, \ and\ \bibinfo {author} {\bibfnamefont {G.~N.}\ \bibnamefont
  {Patey}},\ }\href@noop {} {\bibfield  {journal} {\bibinfo  {journal} {J.
  Chem. Phys.}\ }\textbf {\bibinfo {volume} {125}},\ \bibinfo {pages} {204502}
  (\bibinfo {year} {2006})}\BibitemShut {NoStop}%
\bibitem [{\citenamefont {McPhie}\ \emph {et~al.}(2006)\citenamefont {McPhie},
  \citenamefont {Daivis},\ and\ \citenamefont {Snook}}]{McPhie:2006bn}%
  \BibitemOpen
  \bibfield  {author} {\bibinfo {author} {\bibfnamefont {M.~G.}\ \bibnamefont
  {McPhie}}, \bibinfo {author} {\bibfnamefont {P.~J.}\ \bibnamefont {Daivis}},
  \ and\ \bibinfo {author} {\bibfnamefont {I.~K.}\ \bibnamefont {Snook}},\
  }\href@noop {} {\bibfield  {journal} {\bibinfo  {journal} {Phys. Rev. E}\
  }\textbf {\bibinfo {volume} {74}},\ \bibinfo {pages} {011205} (\bibinfo
  {year} {2006})}\BibitemShut {NoStop}%
\bibitem [{\citenamefont {Ould-Kaddour}\ and\ \citenamefont
  {Levesque}(2007)}]{OuldKaddour:2007gv}%
  \BibitemOpen
  \bibfield  {author} {\bibinfo {author} {\bibfnamefont {F.}~\bibnamefont
  {Ould-Kaddour}}\ and\ \bibinfo {author} {\bibfnamefont {D.}~\bibnamefont
  {Levesque}},\ }\href@noop {} {\bibfield  {journal} {\bibinfo  {journal} {J.
  Chem. Phys.}\ }\textbf {\bibinfo {volume} {127}},\ \bibinfo {pages} {154514}
  (\bibinfo {year} {2007})}\BibitemShut {NoStop}%
\bibitem [{\citenamefont {Jung}\ \emph {et~al.}(2017)\citenamefont {Jung},
  \citenamefont {Hanke},\ and\ \citenamefont {Schmid}}]{Jung:2017gn}%
  \BibitemOpen
  \bibfield  {author} {\bibinfo {author} {\bibfnamefont {G.}~\bibnamefont
  {Jung}}, \bibinfo {author} {\bibfnamefont {M.}~\bibnamefont {Hanke}}, \ and\
  \bibinfo {author} {\bibfnamefont {F.}~\bibnamefont {Schmid}},\ }\href@noop {}
  {\bibfield  {journal} {\bibinfo  {journal} {J. Chem. Theory Comput.}\
  }\textbf {\bibinfo {volume} {13}},\ \bibinfo {pages} {2481} (\bibinfo {year}
  {2017})}\BibitemShut {NoStop}%
\bibitem [{\citenamefont {Hansen}\ and\ \citenamefont
  {McDonald}(2013)}]{Hansen:2013uv}%
  \BibitemOpen
  \bibfield  {author} {\bibinfo {author} {\bibfnamefont {J.~P.}\ \bibnamefont
  {Hansen}}\ and\ \bibinfo {author} {\bibfnamefont {I.~R.}\ \bibnamefont
  {McDonald}},\ }\href@noop {} {\emph {\bibinfo {title} {{Theory of Simple
  Liquids}}}},\ \bibinfo {edition} {4th}\ ed.\ (\bibinfo  {publisher} {Academic
  Press},\ \bibinfo {address} {London},\ \bibinfo {year} {2013})\BibitemShut
  {NoStop}%
\bibitem [{\citenamefont {Li}(2009)}]{Li:2009fv}%
  \BibitemOpen
  \bibfield  {author} {\bibinfo {author} {\bibfnamefont {Z.}~\bibnamefont
  {Li}},\ }\href@noop {} {\bibfield  {journal} {\bibinfo  {journal} {Phys. Rev.
  E}\ }\textbf {\bibinfo {volume} {80}},\ \bibinfo {pages} {8} (\bibinfo {year}
  {2009})}\BibitemShut {NoStop}%
\bibitem [{\citenamefont {Morrone}\ \emph {et~al.}(2012)\citenamefont
  {Morrone}, \citenamefont {Li},\ and\ \citenamefont {Berne}}]{Morrone:2012ba}%
  \BibitemOpen
  \bibfield  {author} {\bibinfo {author} {\bibfnamefont {J.~A.}\ \bibnamefont
  {Morrone}}, \bibinfo {author} {\bibfnamefont {J.}~\bibnamefont {Li}}, \ and\
  \bibinfo {author} {\bibfnamefont {B.~J.}\ \bibnamefont {Berne}},\ }\href@noop
  {} {\bibfield  {journal} {\bibinfo  {journal} {J. Phys. Chem. B}\ }\textbf
  {\bibinfo {volume} {116}},\ \bibinfo {pages} {378} (\bibinfo {year}
  {2012})}\BibitemShut {NoStop}%
\bibitem [{\citenamefont {Chakraborty}(2011)}]{Chakraborty:2011jd}%
  \BibitemOpen
  \bibfield  {author} {\bibinfo {author} {\bibfnamefont {D.}~\bibnamefont
  {Chakraborty}},\ }\href@noop {} {\bibfield  {journal} {\bibinfo  {journal}
  {Eur. Phys. J. B}\ }\textbf {\bibinfo {volume} {83}},\ \bibinfo {pages} {375}
  (\bibinfo {year} {2011})}\BibitemShut {NoStop}%
\bibitem [{\citenamefont {Zwanzig}\ and\ \citenamefont
  {Bixon}(1970)}]{Zwanzig:1970kr}%
  \BibitemOpen
  \bibfield  {author} {\bibinfo {author} {\bibfnamefont {R.}~\bibnamefont
  {Zwanzig}}\ and\ \bibinfo {author} {\bibfnamefont {M.}~\bibnamefont
  {Bixon}},\ }\href@noop {} {\bibfield  {journal} {\bibinfo  {journal} {Phys.
  Rev. A}\ }\textbf {\bibinfo {volume} {2}},\ \bibinfo {pages} {2005} (\bibinfo
  {year} {1970})}\BibitemShut {NoStop}%
\bibitem [{\citenamefont {Chow}\ and\ \citenamefont
  {Hermans}(1972)}]{Chow:1972gr}%
  \BibitemOpen
  \bibfield  {author} {\bibinfo {author} {\bibfnamefont {T.~S.}\ \bibnamefont
  {Chow}}\ and\ \bibinfo {author} {\bibfnamefont {J.~J.}\ \bibnamefont
  {Hermans}},\ }\href@noop {} {\bibfield  {journal} {\bibinfo  {journal} {J.
  Chem. Phys.}\ }\textbf {\bibinfo {volume} {56}},\ \bibinfo {pages} {3150}
  (\bibinfo {year} {1972})}\BibitemShut {NoStop}%
\bibitem [{\citenamefont {Chow}\ and\ \citenamefont
  {Hermans}(1973)}]{Chow:1973cm}%
  \BibitemOpen
  \bibfield  {author} {\bibinfo {author} {\bibfnamefont {T.~S.}\ \bibnamefont
  {Chow}}\ and\ \bibinfo {author} {\bibfnamefont {J.~J.}\ \bibnamefont
  {Hermans}},\ }\href@noop {} {\bibfield  {journal} {\bibinfo  {journal}
  {Physica}\ }\textbf {\bibinfo {volume} {65}},\ \bibinfo {pages} {156}
  (\bibinfo {year} {1973})}\BibitemShut {NoStop}%
\bibitem [{\citenamefont {Hauge}\ and\ \citenamefont
  {Martin-L{\"o}f}(1973)}]{Hauge:1973df}%
  \BibitemOpen
  \bibfield  {author} {\bibinfo {author} {\bibfnamefont {E.~H.}\ \bibnamefont
  {Hauge}}\ and\ \bibinfo {author} {\bibfnamefont {A.}~\bibnamefont
  {Martin-L{\"o}f}},\ }\href@noop {} {\bibfield  {journal} {\bibinfo  {journal}
  {J. Stat. Phys.}\ }\textbf {\bibinfo {volume} {7}},\ \bibinfo {pages} {259}
  (\bibinfo {year} {1973})}\BibitemShut {NoStop}%
\bibitem [{\citenamefont {Bedeaux}\ and\ \citenamefont
  {Mazur}(1974)}]{Bedeaux:1974cd}%
  \BibitemOpen
  \bibfield  {author} {\bibinfo {author} {\bibfnamefont {D.}~\bibnamefont
  {Bedeaux}}\ and\ \bibinfo {author} {\bibfnamefont {P.}~\bibnamefont
  {Mazur}},\ }\href@noop {} {\bibfield  {journal} {\bibinfo  {journal}
  {Physica}\ }\textbf {\bibinfo {volume} {78}},\ \bibinfo {pages} {505}
  (\bibinfo {year} {1974})}\BibitemShut {NoStop}%
\bibitem [{\citenamefont {Hinch}(1975)}]{Hinch:1975ba}%
  \BibitemOpen
  \bibfield  {author} {\bibinfo {author} {\bibfnamefont {E.~J.}\ \bibnamefont
  {Hinch}},\ }\href@noop {} {\bibfield  {journal} {\bibinfo  {journal} {J.
  Fluid Mech.}\ }\textbf {\bibinfo {volume} {73}},\ \bibinfo {pages} {499}
  (\bibinfo {year} {1975})}\BibitemShut {NoStop}%
\bibitem [{\citenamefont {Metiu}\ \emph {et~al.}(1977)\citenamefont {Metiu},
  \citenamefont {Oxtoby},\ and\ \citenamefont {Freed}}]{Metiu:1977hd}%
  \BibitemOpen
  \bibfield  {author} {\bibinfo {author} {\bibfnamefont {H.}~\bibnamefont
  {Metiu}}, \bibinfo {author} {\bibfnamefont {D.~W.}\ \bibnamefont {Oxtoby}}, \
  and\ \bibinfo {author} {\bibfnamefont {K.~F.}\ \bibnamefont {Freed}},\
  }\href@noop {} {\bibfield  {journal} {\bibinfo  {journal} {Phys. Rev. A}\
  }\textbf {\bibinfo {volume} {15}},\ \bibinfo {pages} {361} (\bibinfo {year}
  {1977})}\BibitemShut {NoStop}%
\bibitem [{\citenamefont {Espa{\~n}ol}(1995)}]{Espanol:1995vv}%
  \BibitemOpen
  \bibfield  {author} {\bibinfo {author} {\bibfnamefont {P.}~\bibnamefont
  {Espa{\~n}ol}},\ }\href@noop {} {\bibfield  {journal} {\bibinfo  {journal}
  {Physica A}\ }\textbf {\bibinfo {volume} {214}},\ \bibinfo {pages} {185}
  (\bibinfo {year} {1995})}\BibitemShut {NoStop}%
\bibitem [{\citenamefont {Felderhof}(2005)}]{Felderhof:2005da}%
  \BibitemOpen
  \bibfield  {author} {\bibinfo {author} {\bibfnamefont {B.~U.}\ \bibnamefont
  {Felderhof}},\ }\href@noop {} {\bibfield  {journal} {\bibinfo  {journal} {J.
  Chem. Phys.}\ }\textbf {\bibinfo {volume} {123}},\ \bibinfo {pages} {044902}
  (\bibinfo {year} {2005})}\BibitemShut {NoStop}%
\bibitem [{\citenamefont {Landau}\ and\ \citenamefont
  {Lifshitz}(1987)}]{Landau:1987hg}%
  \BibitemOpen
  \bibfield  {author} {\bibinfo {author} {\bibfnamefont {L.~D.}\ \bibnamefont
  {Landau}}\ and\ \bibinfo {author} {\bibfnamefont {E.~M.}\ \bibnamefont
  {Lifshitz}},\ }\href@noop {} {\emph {\bibinfo {title} {{Fluid Mechanics}}}},\
  \bibinfo {edition} {2nd}\ ed.\ (\bibinfo  {publisher} {Pergamon Press},\
  \bibinfo {address} {Oxford},\ \bibinfo {year} {1987})\BibitemShut {NoStop}%
\bibitem [{\citenamefont {Paul}\ and\ \citenamefont
  {Pusey}(1999)}]{Paul:1981cc}%
  \BibitemOpen
  \bibfield  {author} {\bibinfo {author} {\bibfnamefont {G.~L.}\ \bibnamefont
  {Paul}}\ and\ \bibinfo {author} {\bibfnamefont {P.~N.}\ \bibnamefont
  {Pusey}},\ }\href@noop {} {\bibfield  {journal} {\bibinfo  {journal} {J.
  Phys. A}\ }\textbf {\bibinfo {volume} {14}},\ \bibinfo {pages} {3301}
  (\bibinfo {year} {1999})}\BibitemShut {NoStop}%
\bibitem [{\citenamefont {Zwanzig}\ and\ \citenamefont
  {Bixon}(1975)}]{Zwanzig:1975hk}%
  \BibitemOpen
  \bibfield  {author} {\bibinfo {author} {\bibfnamefont {R.}~\bibnamefont
  {Zwanzig}}\ and\ \bibinfo {author} {\bibfnamefont {M.}~\bibnamefont
  {Bixon}},\ }\href@noop {} {\bibfield  {journal} {\bibinfo  {journal} {J.
  Fluid Mech.}\ }\textbf {\bibinfo {volume} {69}},\ \bibinfo {pages} {21}
  (\bibinfo {year} {1975})}\BibitemShut {NoStop}%
\bibitem [{\citenamefont {Belushkin}\ \emph {et~al.}(2011)\citenamefont
  {Belushkin}, \citenamefont {Winkler},\ and\ \citenamefont
  {Foffi}}]{Belushkin:2011hr}%
  \BibitemOpen
  \bibfield  {author} {\bibinfo {author} {\bibfnamefont {M.}~\bibnamefont
  {Belushkin}}, \bibinfo {author} {\bibfnamefont {R.~G.}\ \bibnamefont
  {Winkler}}, \ and\ \bibinfo {author} {\bibfnamefont {G.}~\bibnamefont
  {Foffi}},\ }\href@noop {} {\bibfield  {journal} {\bibinfo  {journal} {J.
  Phys. Chem. B}\ }\textbf {\bibinfo {volume} {115}},\ \bibinfo {pages} {14263}
  (\bibinfo {year} {2011})}\BibitemShut {NoStop}%
\bibitem [{\citenamefont {Hess}\ \emph {et~al.}(2008)\citenamefont {Hess},
  \citenamefont {Kutzner}, \citenamefont {van~der Spoel},\ and\ \citenamefont
  {Lindahl}}]{Hess:2008db}%
  \BibitemOpen
  \bibfield  {author} {\bibinfo {author} {\bibfnamefont {B.}~\bibnamefont
  {Hess}}, \bibinfo {author} {\bibfnamefont {C.}~\bibnamefont {Kutzner}},
  \bibinfo {author} {\bibfnamefont {D.}~\bibnamefont {van~der Spoel}}, \ and\
  \bibinfo {author} {\bibfnamefont {E.}~\bibnamefont {Lindahl}},\ }\href@noop
  {} {\bibfield  {journal} {\bibinfo  {journal} {J. Chem. Theory Comput.}\
  }\textbf {\bibinfo {volume} {4}},\ \bibinfo {pages} {435} (\bibinfo {year}
  {2008})}\BibitemShut {NoStop}%
\bibitem [{\citenamefont {Abraham}\ \emph {et~al.}(2015)\citenamefont
  {Abraham}, \citenamefont {Murtola}, \citenamefont {Schulz}, \citenamefont
  {P{\'a}ll}, \citenamefont {Smith}, \citenamefont {Hess},\ and\ \citenamefont
  {Lindahl}}]{Abraham:2015gj}%
  \BibitemOpen
  \bibfield  {author} {\bibinfo {author} {\bibfnamefont {M.~J.}\ \bibnamefont
  {Abraham}}, \bibinfo {author} {\bibfnamefont {T.}~\bibnamefont {Murtola}},
  \bibinfo {author} {\bibfnamefont {R.}~\bibnamefont {Schulz}}, \bibinfo
  {author} {\bibfnamefont {S.}~\bibnamefont {P{\'a}ll}}, \bibinfo {author}
  {\bibfnamefont {J.~C.}\ \bibnamefont {Smith}}, \bibinfo {author}
  {\bibfnamefont {B.}~\bibnamefont {Hess}}, \ and\ \bibinfo {author}
  {\bibfnamefont {E.}~\bibnamefont {Lindahl}},\ }\href@noop {} {\bibfield
  {journal} {\bibinfo  {journal} {SoftwareX}\ }\textbf {\bibinfo {volume}
  {1-2}},\ \bibinfo {pages} {19} (\bibinfo {year} {2015})}\BibitemShut
  {NoStop}%
\bibitem [{\citenamefont {Ishii}\ and\ \citenamefont
  {Ohtori}(2016)}]{Ishii:2016bo}%
  \BibitemOpen
  \bibfield  {author} {\bibinfo {author} {\bibfnamefont {Y.}~\bibnamefont
  {Ishii}}\ and\ \bibinfo {author} {\bibfnamefont {N.}~\bibnamefont {Ohtori}},\
  }\href@noop {} {\bibfield  {journal} {\bibinfo  {journal} {Phys. Rev. E}\
  }\textbf {\bibinfo {volume} {93}},\ \bibinfo {pages} {050104} (\bibinfo
  {year} {2016})}\BibitemShut {NoStop}%
\bibitem [{\citenamefont {Tomanek}(2014)}]{Tomanek:2014ba}%
  \BibitemOpen
  \bibfield  {author} {\bibinfo {author} {\bibfnamefont {D.}~\bibnamefont
  {Tomanek}},\ }\href@noop {} {\emph {\bibinfo {title} {{Guide Through the
  Nanocarbon Jungle}}}}\ (\bibinfo  {publisher} {Morgan {\&} Claypool
  Publishers},\ \bibinfo {year} {2014})\BibitemShut {NoStop}%
\bibitem [{\citenamefont {Yeh}\ and\ \citenamefont
  {Hummer}(2004)}]{Yeh:2004gs}%
  \BibitemOpen
  \bibfield  {author} {\bibinfo {author} {\bibfnamefont {I.-C.}\ \bibnamefont
  {Yeh}}\ and\ \bibinfo {author} {\bibfnamefont {G.}~\bibnamefont {Hummer}},\
  }\href@noop {} {\bibfield  {journal} {\bibinfo  {journal} {J. Phys. Chem. B}\
  }\textbf {\bibinfo {volume} {108}},\ \bibinfo {pages} {15873} (\bibinfo
  {year} {2004})}\BibitemShut {NoStop}%
\bibitem [{\citenamefont {Benett}(1976)}]{Benett:1976gj}%
  \BibitemOpen
  \bibfield  {author} {\bibinfo {author} {\bibfnamefont {C.~H.}\ \bibnamefont
  {Benett}},\ }\href@noop {} {\bibfield  {journal} {\bibinfo  {journal} {J.
  Comput. Phys.}\ }\textbf {\bibinfo {volume} {22}},\ \bibinfo {pages} {245}
  (\bibinfo {year} {1976})}\BibitemShut {NoStop}%
\bibitem [{\citenamefont {Subramanian}\ and\ \citenamefont
  {Davis}(1975)}]{Subramanian:1975kv}%
  \BibitemOpen
  \bibfield  {author} {\bibinfo {author} {\bibfnamefont {G.}~\bibnamefont
  {Subramanian}}\ and\ \bibinfo {author} {\bibfnamefont {H.~T.}\ \bibnamefont
  {Davis}},\ }\href@noop {} {\bibfield  {journal} {\bibinfo  {journal} {Phys.
  Rev. A}\ }\textbf {\bibinfo {volume} {11}},\ \bibinfo {pages} {1430}
  (\bibinfo {year} {1975})}\BibitemShut {NoStop}%
\bibitem [{\citenamefont {Grimm}\ \emph {et~al.}(2011)\citenamefont {Grimm},
  \citenamefont {Jeney},\ and\ \citenamefont {Franosch}}]{Grimm:2011cu}%
  \BibitemOpen
  \bibfield  {author} {\bibinfo {author} {\bibfnamefont {M.}~\bibnamefont
  {Grimm}}, \bibinfo {author} {\bibfnamefont {S.}~\bibnamefont {Jeney}}, \ and\
  \bibinfo {author} {\bibfnamefont {T.}~\bibnamefont {Franosch}},\ }\href@noop
  {} {\bibfield  {journal} {\bibinfo  {journal} {Soft Matter}\ }\textbf
  {\bibinfo {volume} {7}},\ \bibinfo {pages} {2076} (\bibinfo {year}
  {2011})}\BibitemShut {NoStop}%
\bibitem [{\citenamefont {Puertas}\ and\ \citenamefont
  {Voigtmann}(2014)}]{Puertas:2014ja}%
  \BibitemOpen
  \bibfield  {author} {\bibinfo {author} {\bibfnamefont {A.~M.}\ \bibnamefont
  {Puertas}}\ and\ \bibinfo {author} {\bibfnamefont {T.}~\bibnamefont
  {Voigtmann}},\ }\href@noop {} {\bibfield  {journal} {\bibinfo  {journal} {J.
  Phys.: Condens. Matter}\ }\textbf {\bibinfo {volume} {26}},\ \bibinfo {pages}
  {243101} (\bibinfo {year} {2014})}\BibitemShut {NoStop}%
\end{thebibliography}
\end{document}